\documentclass[12pt]{article}
\pdfoutput=1

\usepackage{draft} 
\usepackage{hyperref}
\usepackage{graphicx,color}
\usepackage{cite}
\usepackage{mciteplus}
\usepackage{skak}
\usepackage{youngtab}
\DeclareFontFamily{OT1}{pzc}{}
\DeclareFontShape{OT1}{pzc}{m}{it}{<-> s * [1.10] pzcmi7t}{}
\DeclareMathAlphabet{\mathpzc}{OT1}{pzc}{m}{it}

\begin{document}

\unitlength = .8mm

\begin{titlepage}
\rightline{MIT-CTP/4710}

\begin{center}

\hfill \\
\hfill \\
\vskip 1cm

\title{Supersymmetry Constraints and String Theory on K3}

\author{Ying-Hsuan Lin$^\textsymrook$, Shu-Heng Shao$^\textsymrook$, Yifan Wang$^\textsymknight$, Xi Yin$^\textsymrook$}

\address{
$^\textsymrook$Jefferson Physical Laboratory, Harvard University, \\
Cambridge, MA 02138 USA
\\
$^\textsymknight$Center for Theoretical Physics, Massachusetts Institute of Technology, \\
Cambridge, MA 02139 USA}

\email{yhlin@physics.harvard.edu, shshao@physics.harvard.edu, yifanw@mit.edu,
xiyin@fas.harvard.edu}

\end{center}

\abstract{ We study supervertices in six dimensional $(2,0)$ supergravity theories, and derive supersymmetry non-renormalization conditions on the 4- and 6-derivative four-point couplings of tensor multiplets. As an application, we obtain exact non-perturbative results of such effective couplings in type IIB string theory compactified on K3 surface, extending previous work on type II/heterotic duality. The weak coupling limit thereof, in particular, gives certain integrated four-point functions of half-BPS operators in the nonlinear sigma model on K3 surface, that depend nontrivially on the moduli, and capture worldsheet instanton contributions. }

\vfill

\end{titlepage}

\eject

\tableofcontents

\section{Introduction}

Supersymmetry plays an important role in constraining the dynamics of string theory. In toroidal compactifications of type II string theory, up to 14-derivative order couplings in the quantum effective action can be determined as exact functions of the moduli (including the string coupling), by combining supersymmetry non-renormalization conditions and U-duality \cite{Green:1997tv, Green:1998by, Green:2006gt, Green:2010wi, Green:2014yxa, Bossard:2014aea, Bossard:2015uga, Bossard:2015oxa, Wang:2015jna, Wang:2015aua}. In this paper, we extend such results to 4- and 6-derivative order couplings of tensor multiplets in the compactification of type IIB string theory on K3 surface.

We will begin by classifying the supervertices (local S-matrix elements that obey supersymmetry Ward identities) in a 6$d$ $(2,0)$ supergravity theory, at the relevant derivative orders. We will focus on the 4 and 6-derivative couplings of tensor multiplets, of the schematic form 
$$
f^{(4)}_{abcd}(\phi) H^a H^b H^c H^d~~~{\rm  and} ~~~ f^{(6)}_{ab,cd}(\phi) D^2(H^a H^b) H^c H^d,
$$
where $\phi$ stands for the massless scalar moduli fields, that parameterize the moduli space \cite{Aspinwall:1996mn}
\ie
{\cal M} = {\cal O}(\Gamma_{21,5})\backslash SO(21,5)/(SO(21)\times SO(5)),
\label{IIBmodulispace}
\fe
and we have omitted the contraction of the Lorentz indices on the self-dual tensor fields $H^a$ in the tensor multiplets (not to be confused with the anti-self-dual tensor fields in the supergravity multiplet), $a=1,\cdots,21$.

By consideration of the factorization of six-point superamplitudes through graviton and tensor poles, we derive second order differential equations that constrain $f^{(4)}_{abcd}(\phi)$ and $f^{(6)}_{ab,cd}(\phi)$. These equations are of the schematic form
\ie\label{eqnss}
& \nabla_a\cdot\nabla_b f^{(4)} \sim f^{(4)},
\\
& \nabla_a\cdot\nabla_b f^{(6)} \sim f^{(6)} + (f^{(4)})^2.
\fe
By consideration of the duality between type II string theory on $K3\times S^1$ and the heterotic string on $T^5$, we find that $f^{(4)}$ and $f^{(6)}$ are given {\it exactly} by the low energy limit of the one-loop and two-loop heterotic string amplitudes, with the results
\ie\label{ffss}
& f^{(4)}_{abcd} = \left. {\partial^4\over \partial y^a \partial y^b \partial y^c \partial y^d}\right|_{y=0} \int_{{\cal F}_1} {d^2\tau} {\tau_2^{1\over 2}\Theta_\Lambda(y|\tau,\bar\tau)\over \Delta(\tau)},
\\
& f^{(6)}_{ab,cd} = {\epsilon_{IK}\epsilon_{JL}+\epsilon_{IL}\epsilon_{JK}\over 3} \left. {\partial^4\over \partial y^a_I \partial y^b_J \partial y^c_K \partial y^d_L}\right|_{y=0} \int_{{\cal F}_2} {\prod_{I\leq J} d^2\Omega_{IJ}} {\Theta_\Lambda(y|\Omega,\bar\Omega)\over (\det{\rm Im}\Omega)^{1\over 2}\Psi_{10}(\Omega)}.
\fe
Here ${\cal F}_1$ is the fundamental domain of the $SL(2,\mathbb{Z})$ action on the upper half plane, parameterized by $\tau$, and ${\cal F}_2$ is the fundamental domain of the $Sp(4,\mathbb{Z})$ action on the Siegel upper half space, parameterized by the period matrix $\Omega_{IJ}$. $\Theta_\Lambda(y|\tau,\bar\tau)$ is the theta function of the even unimodular lattice $\Lambda$ of signature $(21,5)$, embedded in $\mathbb{R}^{21,5}$, and $\Theta_\Lambda(y|\Omega,\bar\Omega)$ is an analogous genus two theta function. The precise expressions of these theta functions will be given later. The above two expressions  depend on the embedding of the lattice $\Lambda$ into $\mathbb{R}^{21,5}$ through the theta functions, and the space of inequivalent  embeddings is the same as the moduli space $\mathcal{M}$ \eqref{IIBmodulispace} of the 6$d$ $(2,0)$ supergravity.  $\Delta(\tau)=\eta^{24}(\tau)$ is the weight 12 cusp form of $SL(2,\mathbb{Z})$, and $\Psi_{10}(\Omega)$ is the weight 10 Igusa cusp form of $Sp(4,\mathbb{Z})$. The result for the 4-derivative term $f^{(4)}$ has previously been obtained in \cite{Kiritsis:2000zi}.

We will verify, through rather lengthy calculations, that  (\ref{ffss}) indeed obey second order differential equations of the form (\ref{eqnss}), and fix the precise numerical coefficients in these equations.

While the expressions (\ref{ffss}) for the coupling coefficients $f^{(4)}$ and $f^{(6)}$ are fully non-perturbative in type IIB string theory, the results are nontrivial even at string tree-level. For instance, in the limit of weak IIB string coupling $g_{\text{\tiny IIB}}$, $f^{(4)}$ reduces to
\ie
f^{(4)}_{ijk\ell} \to {\sqrt{V_{K3}}\over g_{\text{\tiny IIB}}\ell_s^4} A_{ijk\ell}(\varphi),
\fe
where $\varphi$ denotes collectively $\varphi_i^{\pm\pm}$ ($i=1,\cdots,20$), the moduli of the 2$d$ $(4,4)$ CFT given by the supersymmetric nonlinear sigma model on K3 (we will refer to this as the K3 CFT). From the point of view of the worldsheet CFT, we can express $A_{ijk\ell}(\varphi)$ as an integrated four-point function of marginal BPS operators of the K3 CFT, through the expansion
\ie
& \int {d^2 z\over 2\pi} \,|z|^{-s-1} |1-z|^{-t-1} \left\langle \phi_i^{RR}(z) \phi_j^{RR}(0) \phi_k^{RR}(1) \phi_\ell^{RR}(\infty)\right\rangle
\\
&= {\delta_{ij}\delta_{k\ell}\over s} + {\delta_{ik}\delta_{j\ell}\over t} + {\delta_{i\ell}\delta_{jk}\over u} + A_{ijk\ell} + B_{ij,k\ell} s + B_{ik,j\ell} t + B_{i\ell,jk} u + {\cal O}(s^2,t^2,u^2).
\fe
Here $u=-s-t$, and $\phi_i^{RR}(z)$ are the weight $({1\over 4},{1\over 4})$ RR sector superconformal primaries in the R-symmetry singlet, related to the NS-NS sector weight $({1\over 2},{1\over 2})$, exactly marginal, superconformal primaries by spectral flow. The $z$-integral is defined using Gamma function regularization, or equivalently, analytic continuation in $s$ and $t$ from the domain where the integral converges. While $A_{ijk\ell}$ gives the tree-level contribution to $f^{(4)}$, $B_{ij,k\ell}$ captures the tree-level contribution to $f^{(6)}$.

Note that, in contrast to the Riemannian curvature of the Zamolodchikov metric \cite{Zamolodchikov:1986gt}, which is contained in a contact term of the four-point function \cite{Kutasov:1988xb}, $A_{ijk\ell}$ and $B_{ij,k\ell}$ are determined by the non-local part of the four-point function and do not involve the contact term. Unlike the Zamolodchikov metric which has constant curvature on the moduli space of K3 (with the exception of orbifold type singularities), $A_{ijk\ell}$ and $B_{ij,k\ell}$ are nontrivial functions of the moduli. In particular, the latter coefficients blow up at the points of the moduli space where the CFT becomes singular, corresponding to the K3 surface developing an ADE type singularity, with zero $B$-field flux through the exceptional divisors.

We can give a simple formula for $A_{ijk\ell}$ in the case of $A_1$ ALE target space, which may be viewed as a certain large volume limit of the K3. In this case, the indices $i,j,k,\ell$ only take a single value (denoted by 1), corresponding to a single multiplet that parameterizes the 4-dimensional moduli space
\ie
{\cal M}_{A_1} = {\mathbb{R}^3\times S^1\over \mathbb{Z}_2}.
\fe
${\cal M}_{A_1}$ has two orbifold fixed points by the $\mathbb{Z}_2$ quotient, one of which corresponds to the $\mathbb{C}^2/\mathbb{Z}_2$ free orbifold CFT, whereas the other corresponds to a singular CFT, singular in the sense of a continuous spectrum, that is described by the ${\cal N}=4$ $A_1$ cigar CFT \cite{Ooguri:1995wj,Kutasov:1995te,Giveon:1999px}. While the Zamolodchikov metric does not exhibit any distinct feature between these two points on the moduli space, the integrated four-point function $A_{1111}$ does. The latter is a harmonic function on ${\cal M}_{A_1}$, is finite at the free orbifold point, but blows up at the $A_1$ cigar point. When the $A_1$ singularity is resolved, in the limit of large area of the exceptional divisor, we find that $A_{1111}$ receives a one-loop contribution in $\A'$, plus worldsheet instanton contributions \eqref{instanton}.

The paper is organized as follows. In Section \ref{sec:supervertex} we set up the super-spinor-helicity formalism in 6$d$ $(2,0)$ supergravity and classify the supervertices of low derivative orders. In Section \ref{sec:diff}, we derive the differential equation constraints on the four-point 4- and 6-derivative coupling between the tensor multiplets based on the absence of certain six-point supervertices, with some model-independent constant coefficients yet to be determined. In Section \ref{sec:example}, using type II/heterotic duality, we obtain the exact non-perturbative 4- and 6-derivative couplings in type IIB string theory on K3. We verify that these couplings indeed satisfy the differential equations and fix the constant coefficients in these equations. In Section \ref{sec:K3}, we consider the weak coupling limit of the above results, which gives the integrated four-point function of BPS primaries in the K3 CFT, with an explicit dependence on the moduli space. We also consider the $A_1$ ALE sigma model limit of the K3 CFT and study the 4-derivative couplings in that limit.

\section{Supervertices in 6$d$ $(2,0)$ supergravity}\label{sec:supervertex}

\subsection{6$d$ $(2,0)$ Super-spinor-helicity formalism}

Following \cite{Dennen:2009vk, Cheung:2009dc, Boels:2012ie}, we adopt the convention for 6$d$ spinor-helicity variables
\ie
p_{AB} = \zeta_{A\A} \zeta_{B\B}\epsilon^{\A\B},~~~~ p^{AB}\equiv {1\over 2} \epsilon^{ABCD} p_{CD} = \widetilde\zeta^A{}_{\dot \A} \widetilde\zeta^B{}_{\dot \B} \epsilon^{\dot \A\dot \B},
\fe
and define Grassmannian variables $\eta_{\A I}$ and $\widetilde\eta_{\dot \A I}$, where the lower and upper $A,B$ are $SO(5,1)$ chiral and anti-chiral spinor indices respectively,  $(\A,\dot \A)$ are $SU(2)\times SU(2)$ little group indices, and $I=1,2$ is an auxiliary index which may be identified with the spinor index of an $SO(3)$ subgroup of the $SO(5)$ R-symmetry group. 

Let us represent the 1-particle states in the $(2,0)$ tensor multiplet and the $(2,0)$ supergravity multiplet as polynomials in the Grassmannian variables $\eta_{\A I}$ and $\tilde \eta_{\dot \A I}$. The 1-particle states of the $(2,0)$ tensor multiplet transform in the following representations of the $SU(2)\times SU(2)$ little group,
\ie
{\bf (3,1)} \oplus 4 {\bf(2,1)} \oplus 5{\bf(1,1)}\,.
\fe
 These 1-particle states can be represented collectively as a polynomial 
 \ie
 P(\eta)
 \fe
up to degree 4 in $\eta$, but with no $\tilde \eta$. In particular, the monomial $\eta_{\A I}\eta_{\B J}\epsilon^{IJ}$ corresponds to the self-dual two form $\bf(3,1)$ and the monomials $1,\, \eta_{\A I} \eta_{\B J} \epsilon^{\A\B},\, \eta^4$ correspond to the 5 scalars $\bf(1,1)$.

  The 1-particle states of the $(2,0)$ supergravity multiplet, on the other hand, transform in the following representations of the $SU(2)\times SU(2)$ little group,
  \ie
{ \bf  (3,3)}\oplus 4{\bf (2,3)} \oplus 5{\bf (1,3)}\,.
  \fe
These states are represented by 
\ie
P(\eta) \widetilde\eta_{\dot \A I}\widetilde\eta_{\dot \B J} \epsilon^{IJ}.
\fe
In particular, the monomial $P(\eta)=\eta_{\A I}\eta_{\B J}\epsilon^{IJ}$ corresponds to the graviton $\bf (3,3)$ and the monomials $P(\eta)=1$, $(\eta^2)_{IJ}\equiv\eta_{\A I} \eta_{\B J}\epsilon^{\A\B}$, and $\eta^4$ correspond to the 5 anti-self-dual tensor fields $\bf (1,3)$.

The 16 supercharges are represented on 1-particle states as
\ie
q_{A I} = \zeta_{A \A} \eta^\A{}_I,~~~ \overline{q}_{A I} = \zeta_{A \A} {\partial\over \partial \eta_\A{}^I}.
\fe
They obey the supersymmetry algebra
\ie
\{ q_{AI}, \overline{q}_{BJ}\} = p_{AB} \epsilon_{IJ},~~~~\{q,q\} = \{\overline q,\overline q\}=0.
\fe
The 10 $SO(5)$ R-symmetry generators are
\ie
(\eta^2)_{IJ},\quad (\partial_\eta^2)_{IJ},\quad \eta_I \partial_{\eta_J}-\D^J_I
\label{SO5R}
\fe
when acting on 1-particle states.

In an $n$-point scattering amplitude, we will associate to each particle spinor helicity variables $\zeta_{i A \A}, \, \tilde \zeta_{i A \dot\A}$ and Grassmannian variables $\eta_{i \A I},\, \tilde \eta_{i \dot\A I}$, with $i=1,\cdots, n$. Correspondingly we define the supercharges for each particle,
\ie
q_{i A I} = \zeta_{iA \A} \eta^\A{}_{iI},~~~ \overline{q}_{iA I} = \zeta_{iA \A} {\partial\over \partial \eta_{i\A}{}^I}.
\fe
The supercharges acting on the amplitude are represented by sums of the 1-particle representations
\ie
Q_{AI} = \sum_i q_{iAI}, \quad \overline Q_{AI} = \sum_i \overline q_{iAI},
\fe
and so are the R-symmetry generators
\ie
\sum_i (\eta_i^2)_{IJ}, \quad \sum_i (\partial_{\eta_i}^2)_{IJ}, \quad \sum_i \eta_{iI} \partial_{\eta_{iJ}} - \D^J_I.
\fe

The solutions to the supersymmetry Ward identities can be expressed in terms of the super-spinor-helicity variables. If such expression is local in these variable, we call it a \textit{supervertex}, otherwise it is a \textit{superamplitude}. Among all the supervertices, the {\it D-term} type takes the form
\ie
\D^8(Q)\overline Q^8 \cP(\zeta_i,\tilde \zeta_i, \eta_i,\tilde\eta_i),
\fe
where $\D^8(Q) = \prod_{A,I}Q_{AI}$, and $\cP$ is a polynomial in the super-spinor-helicity variables $\zeta_i,\tilde \zeta_i,\eta_i,\tilde\eta_i$ associated with the external particles labeled by $i=1,\dots, n$, that is Lorentz invariant and little group invariant. On the other hand, the {\it F-term} supervertices are of the form
\ie
\D^8(Q) {\cal F}(\zeta_i,\tilde \zeta_i, \eta_i,\tilde\eta_i),
\fe
where $\cal F$ is a Lorentz invariant and little group invariant polynomial in the super-spinor-helicity variables that cannot be written in the D-term form \cite{Elvang:2009wd, Elvang:2010jv, Wang:2015jna, Lin:2015ixa, Chen:2015hpa}. From momentum counting, we expect D-term supervertices in general to come at or above 8-derivative order.

In the following subsections, we will focus on three- and four-point supervertices in the $(2,0)$ supergravity. We will start with supervertices involving tensor multiplets only, whose classification coincides with that of the $(2,0)$ SCFT on the tensor branch. We will then introduce couplings to the supergravity multiplet and classify the supervertices thereof. In particular, we will discover that the four-point D-term supervertices involving supergravitons do not appear until at 12-derivative order.

\subsection{Supervertices for tensor multiplets}
Among the four-point supervertices that only involve the $(2,0)$ tensor multiplets, the leading F-term ones arise at $4$ and $6$-derivative orders and take the form
\ie
&\D^8(Q)f^{(4)}_{abcd},\\
&\D^8(Q)( f^{(6)}_{ab,cd} s +f^{(6)}_{ac,bd} t + f^{(6)}_{ad,bc} u),
\fe
where $Q_{AI}=\sum_{i=1}^4 q_{iAI}$ and $\D^8(Q) = \prod_{A,I}Q_{AI}$. The coefficients $f^{(4)}, f^{(6)}$ are constant in $s,t,u$ but functions of the moduli. Their dependence on the moduli is the main object of the current paper. The subscripts $a,b,c,d$ label the 21 tensor multiplets. They contain the $H^4$ and $D^2 H^4$ couplings, respectively, where $H$ denotes the self-dual three form field strength in the 21 tensor multiplets.

There are also four-point D-term supervertices of the form $\D^8(Q) \overline Q^8 \cP(\zeta_i,\tilde \zeta_i, \eta_i)$.  For this expression to be non-vanishing, we need $\cP$ to contain at least eight $\eta$'s.  On the other hand, by exchanging the order of $\D^8(Q)$ and $\overline Q^8$, we see that we cannot have more than eight $\eta$'s in $\cP$ because there are in total $4 \times 4$ $\eta$'s from the four 1-particle states.  Hence the lowest derivative order D-term supervertices for tensor multiplets arise at $8$-derivative order
\ie
\D^8(Q) \overline Q^8 \sum_{i<j}\eta_i^4 \eta_j^4 
\fe
This is the unique D-term supervertex of tensor multiplets at 8-derivative order.  Although we could act $\overline{Q}$ on other little group singlets made out of eight $\eta_i$'s, like for instance $(\eta_1^2)_{IJ} (\eta_2^2)^{IJ} (\eta_3^2)_{KL} (\eta_4^2)^{KL}$, such expressions always turn out to be proportional to $\sum_{i<j}\eta_i^4 \eta_j^4$.

Next, we will show that three-point supervertices of tensor multiplets are absent. 
In general it is more intricate to write down the three-point supervertices due to the kinematic constraints,\footnote{ As will be shown in this section, for any choice of the three momenta, two $Q_{AI}$'s and two $\overline Q_{AI}$'s vanish.  While this implies that $\D^8(Q)=0$ (and hence the naive construction of the supervertices as in the four-point and higher-point cases does not apply), the supersymmetry Ward identities associated with the two vanishing $Q_{AI}$'s also become trivial, which means that the full factor of $\D^8(Q)$ is not needed in a supervertex. 
} and we will work in a frame where the three momenta $p_1, p_2, p_3$ lie in a null plane spanned by
 $ e^0+e^1$ (the $0^-$-direction) and $ e^2+ie^3$ (the $1^-$-direction). 
The null plane is equivalently specified by the linear operator,
\ie
\widehat N = p_1^m p_2^n \Gamma_{mn} \equiv (p^+)^2 \Gamma^{0^-1^-},
\fe
such that the spinor helicity variables associated with the momenta satisfy
\ie\label{nz}
\widehat N_A{}^B \zeta_{iB\A}=0,~~~~\widehat N^A{}_B \widetilde\zeta_i{}^B{}_{\dot \A} = 0.
\fe
We write both the lower (chiral) and upper (anti-chiral) $SO(5,1)$ spinor index $A$ as $(\pm\pm)$ which represent spins on the $01$ and $23$ planes, while the spin in the $45$ plane is fixed by the $01$ and $23$ spins due to the chirality condition.  For instance, we write $\zeta_{iA\A}$ as $\zeta_{i\A}^{\pm\pm}$, and $\widetilde\zeta_i{}^A{}_{\dot \A}$ as $\widetilde\zeta_{i\dot \A}^{\pm\pm}$.  By definition, $\zeta_{i\A}^{s_0 s_1}$ (or $\widetilde\zeta_{i\dot \A}^{s_0 s_1}$) has charge ${s_0 \over 2}$ and ${s_1 \over 2}$ under the $SO(1, 1)_{01}$ boost and $SO(2)_{23}$ rotation in the 01 and 23 planes, respectively.  Then by the chirality condition, $\zeta_{i\A}^{s_0 s_1}$ has charge ${(-1)^{{s_0 + s_1 \over 2}} \over 2}$ and $\widetilde\zeta_{i\A}^{s_0 s_1}$ has charge $ -{(-1)^{{s_0 + s_1 \over 2}} \over 2}$ under the $SO(2)_{45}$ ``tiny group" that rotates the 45 plane.  The momentum $p^+$ has charge ${1 \over 2}$ under both the $SO(1, 1)_{01}$ and $SO(2)_{23}$, and is not charged under the $SO(2)_{45}$.  For clarity, these charges are summarized in Table~\ref{tab:charges}.

\begin{table}
\centering
\begin{tabular}{|c|c|c|c|}
\hline
Symbol & $SO(1, 1)_{01}$ & $SO(2)_{23}$ & $SO(2)_{45}$
\\\hline\hline
$\zeta^{++}$ / $\widetilde\zeta^{++}$ & ${1 \over 2}$ & ${1 \over 2}$ & ${1 \over 2}$ / $-{1 \over 2}$
\\
$\zeta^{+-}$ / $\widetilde\zeta^{+-}$ & ${1 \over 2}$ & $-{1 \over 2}$ & $-{1 \over 2}$ / ${1 \over 2}$
\\
$\zeta^{-+}$ / $\widetilde\zeta^{-+}$ & $-{1 \over 2}$ & ${1 \over 2}$ & $-{1 \over 2}$ / ${1 \over 2}$ 
\\
$\zeta^{--}$ / $\widetilde\zeta^{--}$ & $-{1 \over 2}$ & $-{1 \over 2}$ & ${1 \over 2}$ / $-{1 \over 2}$
\\
$p^+$ & ${1 \over 2}$ & ${1 \over 2}$ & 0
\\\hline
$p_1^+$ & 1 & 0 & 0
\\
$p_2^+$ & 0 & 1 & 0
\\
\hline
\end{tabular}
\caption{ The charges of different symbols ($SO(1, 5)$ representations) under the boost and rotations in the three orthogonal planes.  Note that in the last two rows, we choose a frame where $p_1$ is parallel to $e^0 + e^1$, and $p_2$ is parallel to $e^2 + i e^3$. }
\label{tab:charges}
\end{table}


The constraint (\ref{nz}) implies that $\zeta_{i\A}^{--}=\widetilde\zeta_{i\dot \A}^{--}=0$. Consequently the supercharges $Q^{--}_I$ and $\overline{Q}^{--}_I$ vanish identically. The expression
\ie
\prod_{I=1,2} Q_I^{+-}Q_I^{-+}Q_I^{++}
\label{Q6}
\fe
is thus annihilated by all 16 supercharges $Q_{AI}$ and $\overline{Q}_{AI}$. Since \eqref{Q6} has $SO(2)_{45}$ tiny group charge $-1$, a general three-point supervertex for the tensor multiplets must take the following form
\ie\label{3ptensor}
\left(\prod_{I=1,2} Q_I^{+-}Q_I^{-+}Q_I^{++} \right) f_{abc}(\zeta_i,\eta_i),
\fe 
where $f_{abc}$ must be annihilated by $\overline{Q}$ up to terms proportional to $Q$, invariant with respect to the little groups, and have charge $+1$ under tiny group. By consideration of CPT conjugation,\footnote{
For an $n$-point ($n \geq 4$) supervertex or superamplitude
\ie
{\cal V} = \delta^{8} (Q) {\cal F}(\zeta_i,\eta_i; \widetilde\zeta_i,\widetilde\eta_i),
\fe
the CPT conjugate is
\ie
\overline{\cal V} = \overline Q^{8} {\cal F}(\zeta_i,\partial/\partial\eta_i; \widetilde\zeta_i, \widetilde\eta_i) \prod_{i=1}^n \eta_i^4  .
\fe
For a three-point supervertex
\ie
{\cal V} = \prod_{I=1,2} Q_I^{+-}Q_I^{-+}Q_I^{++} {\cal F}(\zeta_i,\eta_i; \widetilde\zeta_i,\widetilde\eta_i),
\fe
the CPT conjugate is
\ie
\overline{\cal V} = \prod_{I=1,2} \overline{Q}_I^{+-} \overline{Q}_I^{-+} \overline{Q}_I^{++} {\cal F}(\zeta_i,\partial/\partial\eta_i; \widetilde\zeta_i, \widetilde\eta_i) \prod_{i=1}^n \eta_i^4  .
\fe
}
$f_{abc}$ cannot depend on $\eta_i$ (otherwise the CPT conjugate expression would involve fewer than 6 $\eta$'s and cannot be proportional to $Q^6$). Little group invariance then forces it to be a function of the momenta only. In particular, since all three momenta are $SO(2)_{45}$ tiny group invariant, $f_{abc}$ would have to be tiny group invariant by itself which then forces it to vanish.

\subsection{Supervertices for supergravity and tensor multiplets}
We will now incorporate the coupling to the supergravity multiplet. Below to ease the notation, we will define
\ie
\widetilde q^A{}_I \equiv \widetilde\zeta^A{}_{\dot \A}\widetilde \eta^{\dot \A}{}_I,~~~~(\widetilde q^2)^{AB}\equiv \widetilde q^A{}_I\widetilde q^B{}_J \epsilon^{IJ}.
\fe

\paragraph{Four-point supervertices.}  The four-point F-term supervertex of supergravity multiplet arises at $8$-derivative order,
\ie
\delta^8(Q) (\widetilde q_1^2)^{AA'} (\widetilde q_2^2)^{BB'} (\widetilde q_3^2)^{CC'} (\widetilde q_4^2)^{DD'} \epsilon_{ABCD}\epsilon_{A'B'C'D'}.
\fe
which includes the $R^4$ coupling.
The lowest derivative order D-term four-point supervertex is
\ie
& \delta^8(Q)\overline{Q}^8 \sum_{i<j}\eta_i^4\eta_j^4\, (\widetilde q_1^2)^{AA'} (\widetilde q_2^2)^{BB'} (\widetilde q_3^2)^{CC'} (\widetilde q_4^2)^{DD'} \epsilon_{ABCD}\epsilon_{A'B'C'D'}
\\
&=\delta^8(Q) \sum_{i<j} s_{ij}^2 \, (\widetilde q_1^2)^{AA'} (\widetilde q_2^2)^{BB'} (\widetilde q_3^2)^{CC'} (\widetilde q_4^2)^{DD'} \epsilon_{ABCD}\epsilon_{A'B'C'D'},
\fe
which is at 12-derivative order and contains the  $D^4R^4$ coupling.\footnote{ This is also the only D-term supervertex of supergravity multiplet at the 12-derivative order.  The $\widetilde \eta$'s anti-commute with the supercharges. Their only role is to form supergraviton states and they must be contracted with the $\widetilde \zeta$'s to form little group singlets.}

We also have a four-point F-term supervertex at 8-derivative order that involves one tensor multiplet and two supergravity multiplets as external states
\ie
\D^8(Q) (\widetilde q_1^2)^{AB} (\widetilde q_2^2)^{CD}  p_{3AC}p_{4BD}  ,
\label{D2R2H2}
\fe
which contains the $D^2(R^2 H^2)$ coupling.\footnote{ The $6$-derivative order supervertex that contains the $R^2H^2$ coupling appears to be absent.}  We can also obtain a 10-derivative F-term by multiplying the 8-derivative one (\ref{D2R2H2}) by $s_{12}$, which contains the $D^4 (R^2 H^2)$ coupling.  The lowest derivative order D-term is
\ie
\D^8(Q) \overline Q^8 \eta_3^4 \eta_4^4  (\widetilde q_1^2)^{AB} (\widetilde q_2^2)^{CD}  p_{3AC}p_{4BD} ,
\fe
which is at 12-derivative order and contains the $D^6 (R^2 H^2)$ coupling.

The fact that D term four-point supervertices involving the supergravity multiplet only start appearing at 12-derivative order is a special feature of $(2,0)$ supergravity, in contrast to the naive momentum counting that may suggest they occur at 8-derivative order (as in the case of maximally supersymmetric gauge theories, with sixteen supersymmetries).  

\paragraph{Three-point supervertices.}  Let us now discuss the three-point supervertices between the $(2,0)$ supergravity multiplet and the tensor multiplets. Below we will explicitly construct the 2-derivative supervertices and also argue for the absence of three-point supervertices at 4-derivative order and beyond.
 
At 2-derivative order, the 3-supergraviton supervertex is given by
\ie\label{3p2dg}
{1\over (p^+)^4}\left(\prod_{I=1,2} Q_I^{+-}Q_I^{-+}Q_I^{++} \right) (\widetilde q_1^2)^{AA'} (\widetilde q_2^2)^{BB'} (\widetilde q_3^2)^{CC'} \epsilon_{ABC,--}\epsilon_{A'B'C',--}.
\fe 
The power of $p^+$ is fixed by the $SO(1, 1)_{01}$ and $SO(2)_{23}$ invariance, and this expression is also invariant under the $SO(2)_{45}$ tiny group, thereby consistent with the full $SO(1, 5)$ Lorentz symmetry.

More generally, a cubic supervertex of the supergravity multiplet must be of the form
\ie
\left( \prod_{I=1,2} Q_I^{+-}Q_I^{-+}Q_I^{++} \right) (\widetilde q_1^2)^{AA'} (\widetilde q_2^2)^{BB'} (\widetilde q_3^2)^{CC'} P_{ABCA'B'C'}(\zeta_i,\eta_i).
\fe
$P_{ABCA'B'C'}$ must be annihilated by $\overline{Q}$ up to terms proportional to $Q$, invariant with respect to the little groups, and must have charge $2$ under tiny group scaling. As we have argued for the 3-tensor supervertices in the previous subsection, by applying CPT conjugation and little group invariance, we conclude $P_{ABCA'B'C'}$ is a tiny group invariant that only depends the momenta.
 The tiny group invariance of the full amplitude then forces $(AA',BB',CC')$ to have a total of 4 $-$'s and 8 $+$'s, and then $SO(1, 1)_{01}$ and $SO(2)_{23}$ invariance forces $P_{ABCA'B'C'}$ to scale like $(p^+)^{-4}$, and we are back to the two-derivative cubic supervertex \eqref{3p2dg}. This rules out any higher derivative cubic supervertices of the supergravity multiplet.

Now let us consider the three-point supervertex for one supergravity and two tensor multiplets. We can further choose the lightcone coordinates to be aligned with the momenta of the first and second particle, by demanding that $p_1= p_1^+(e^0+e^1)$, and $p_2=p_2^+(e^2+ie^3)$. This amounts to the restriction
\ie
\zeta_1^{-+} =0,~~~~ \zeta_2^{+-}=0.
\fe
At two-derivative order, the gravity-tensor-tensor supervertex is 
\ie
{1\over (p_1^{+})^2}\left(\prod_{I=1,2} Q_I^{+-}Q_I^{-+}Q_I^{++} \right) (\widetilde q_1^2)^{(+-,+-)} ,
\fe
where $p_1$ labels the momentum of the supergraviton.   At 4-derivative order and beyond, there do not appear to be three-point supervertices for the gravity-tensor-tensor coupling, using the same argument as above.  Similarly one can argue that no gravity-gravity-tensor supervertex exists.\footnote{ It appears that one can write down a 2-derivative order supervertex
\ie
{1\over (p_1^{+})^2 p_2^{+}}\left(\prod_{I=1,2} Q_I^{+-}Q_I^{-+}Q_I^{++} \right) (\widetilde q_1^2)^{(+-,+-)}(\widetilde q_2^2)^{(++,-+)}+(1\leftrightarrow 2).
\fe
However, after restoring the full $SO(1, 5)$ Lorentz invariance, the resulting expression cannot be a local supervertex.  This can be seen by noting that the expression is $SO(5)$ R-symmetry invariant, and there simply does not exist any 2-derivative three-point coupling that involves two fields from the gravity multiplet and the self-dual tensor field, which is the only component of the tensor multiplet that is R-symmetry invariant.
}

It is claimed in \cite{Gregori:1997hi} that in type IIB string theory on K3, there is a CP-odd $RH^2$ effective coupling that arises at one-loop order, where here $H$ refers to a mixture of the self-dual two-form in a tensor multiplet and the anti-self-dual two-form in the multiplet that also contains the dilaton . This would seem to correspond to a 4-derivative cubic supervertex. A more careful inspection of the 6$d$ IIB cubic vertex of \cite{Gregori:1997hi} shows that it in fact vanishes identically \cite{Liu:2013dna}, which is consistent with our finding based on the super spinor helicity formalism.

The classification of three-point and four-point supervertices given in this section is summarized in Table~\ref{tab:supvtx}.
\begin{table}[ht]
\centering
\begin{tabular}{|c|c|}
\hline
Supervertices & Derivative Order
\\\hline\hline
$ggg$ & 2 only
\\\hline
$gtt$ & 2 only
\\\hline
$ggt$ & absent
\\\hline
$ttt$ & absent
\\\hline\hline
$gggg$ & F-terms: 8 and possibly 12+
\\
& D-terms: 12+
\\\hline
$ggtt$ & F-terms: 8, 10, and possibly 12+
\\
& D-terms: 12+
\\\hline
$tttt$ & F-terms: 4, 6, and possibly 8+
\\
& D-terms: 8+
\\\hline
\end{tabular}
\caption{Classification of supervertices in 6$d$ $(2,0)$ supergravity.  Here $g$ and $t$ refer to the gravity and tensor supermultiplets which include $R$ and $H$, respectively.  The derivative order includes the derivatives implicit in the fields.  For example, $D^2 (R^2 H^2)$ is regarded as an 8-derivative supervertex ($2 + 2 \times 2 + 2 \times 1$). }
\label{tab:supvtx}
\end{table}
In particular, the three- and four-point supervertices are all \textit{invariant} under the $SO(5)$ R-symmetry \eqref{SO5R}.  In other words, our classification implies that $SO(5)$ breaking supervertices in $(2,0)$ supergravity can only start appearing at five-point and higher.  The simplest examples of such supervertices are $\D^8(Q)$ at $n$-point with $n>4$, which transform in the $[n-4,0]$ representations of the $SO(5)$ R-symmetry~\cite{Cordova:2015vwa}.

\section{Differential constraints on $f^{(4)}$ and $f^{(6)}$ couplings}
\label{sec:diff}

In this section, we shall deduce the general structures of the differential constraints on $f^{(4)}$ and $f^{(6)}$ couplings due to supersymmetry, using superamplitude techniques \cite{Wang:2015jna, Lin:2015ixa, Wang:2015aua}. 

The construction of the $f^{(4)}$ and $f^{(6)}$ supervertices in $(2,0)$ supergravity gives the on-shell supersymmetric completion of the $H^4$ and $D^2H^4$ couplings. In particular, given their relatively low derivative orders, such supervertices must be of F-term type which are rather scarce and have been classified and explicitly constructed in the previous section. As we shall see below, the absence of certain higher point supervertices of these derivative orders will lead to differential constraints on the moduli dependence of the aforementioned couplings in the quantum effective action of $(2,0)$ supergravity.

\begin{figure}[htb]
\centering
\includegraphics[scale=1.5 ]{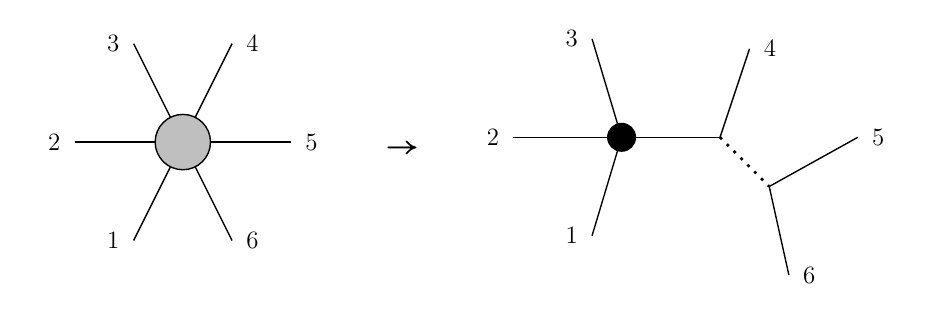}\\
\caption{Factorization channels for the $\varphi^2H^4$ superamplitude. The solid lines stand for the tensor multiplet states while the dotted lines stand for the supergravity multiplet states. The black circles represent the 4-derivative four-tensor-multiplet supervertex, and the trivalent vertices represent the 2-derivative supervertex involving one gravity and two tensor multiplets.}
\label{fig:f4}
\end{figure}

For example, we can expand the supersymmetric $f^{(4)}$ coupling, in terms of the moduli fields, and obtain higher-point vertices. In particular, the resulting six-point $\varphi^2 H^4$ coupling in the singlet representation of $SO(5)$ R-symmetry can be related to a {\it symmetric double soft limit} of the corresponding six-point superamplitude (at 4-derivative order) \cite{Cordova:2015vwa, Wang:2015aua}.
The absence of $SO(5)$ R-symmetry invariant six-point supervertices at 4-derivative order~\cite{Cordova:2015vwa} means that this six-point $\varphi^2 H^4$ coupling from expanding $f^{(4)}$ cannot possibly have a local supersymmetric completion. Rather, it must be related to polar pieces of the superamplitude via supersymmetry; in other words, it is fixed by the residues in all factorization channels. The $\varphi^2 H^4$ superamplitude can only factorize through the 4-derivative supervertex for tensor multiplets and 2-derivative cubic supervertices for two tensor and one graviton multiplets (see Figure~\ref{fig:f4}), giving rise to
\ie
\nabla_{(e}\cdot \nabla_{f)} f^{(4)}_{abcd}=U f^{(4)}_{abcd}\D_{ef}+V f^{(4)}_{(\underline{e}(abc}\D_{d)\underline{f})}+Wf^{(4)}_{ef(ab}\D_{cd)}.
\label{f4diffeqn:gen}
\fe

Here a natural $SO(21,5)$ homogeneous vector bundle $\cW$ over $\cM$ arises as the quotient 
\ie
{SO(21,5)\times {\mathbb R}^{21,5}\over SO(21)\times SO(5)},
\fe
where $\mathbb R^{21,5}$ transforms as a vector under $SO(21)\times SO(5)$. We define the covariant derivative $\nabla_{ai}$, where $a = 1, \dotsc, 21$ and $i = 1, \dotsc, 5$, by the $SO(21,5)$ invariant connection on $\cW$ that gives rise to the symmetric space structure of the scalar manifold $\cM$. Further imposing invariance under the $SO(5)$ R-symmetry means we can focus on the $SO(21)$ subbundle $\cV\subset \cW$. The coupling $f^{(4)}_{abcd}$ becomes a section of the symmetric product vector bundle $\cV_{\tiny \yng(4)}$, on which the second order differential operator $\nabla_{(e}\cdot\nabla_{f)}$ acts naturally.

\begin{figure}[!htb]
\centering
\includegraphics[scale=1.1]{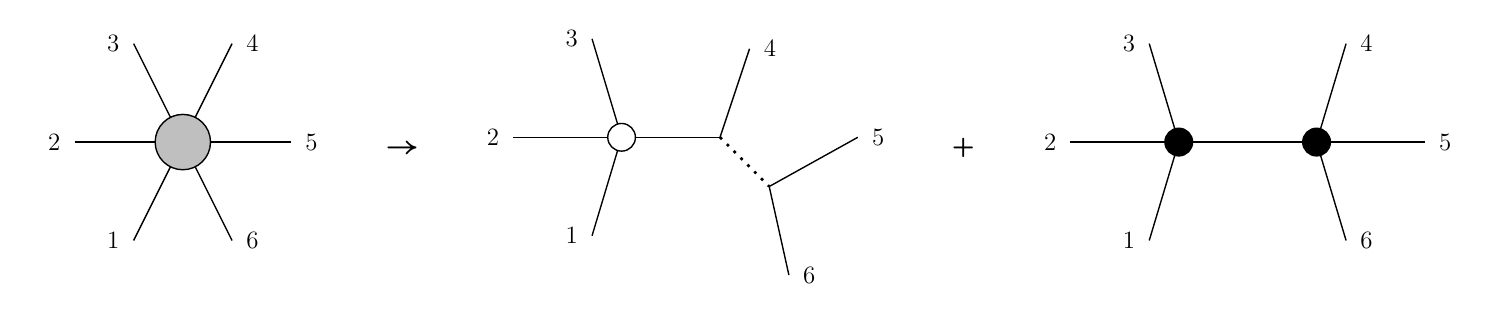}\\
\caption{Factorization channels for the $D^2(\varphi^2H^4)$ superamplitude.  The solid lines stand for the tensor multiplet states while the dotted lines stand for the supergravity multiplet states. The black and white circles represent the 4 and 6-derivative four-tensor-multiplet supervertices, respectively, and the trivalent vertices represent the 2-derivative supervertex involving one gravity and two tensor multiplets.}
\label{fig:f6}
\end{figure}

For the $f^{(6)}$ coupling, recall that it is defined as the coefficient in the superamplitude
\ie
\D^8(Q)( f^{(6)}_{ab,cd} s + f^{(6)}_{ac,bd} t + f^{(6)}_{ad,bc} u).
\fe
Due to the relation $s + t + u = 0$, there is an ambiguity in the definition of $f^{(6)}_{ab, cd}$, where we can shift $f^{(6)}$ by a term that is totally symmetric in three of the four indices.  We fix this ambiguity by demanding that $f^{(6)}_{a(b, cd)} = 0$, which makes $f^{(6)}$ a section of the $\cV_{\tiny\yng(2,2)}$ vector bundle.  The corresponding $D^2(\varphi^2 H^4)$ superamplitude can also factorize through two $f^{(4)}$ supervertices (see Figure~\ref{fig:f6}), and we end up with the following differential constraint
\ie
2\nabla_{(e}\cdot \nabla_{f)} f^{(6)}_{ab,cd}
&=u_1  f^{(6)}_{ab,cd} \delta_{ef}
+u_2 (   f^{(6)}_{ef ,ab}\delta_{cd}  +  f^{(6)}_{ef,cd}\delta_{ab}  )
+u_2'  f^{(6)}_{ef ,(\underline{c}(a} \delta_{b) \underline{d})}\\
&+u_3 ( f^{(6)}_{ea,fb}\delta_{cd} + f^{(6)}_{ec,fd}\delta_{ab})
+u_3'  f^{(6)}_{e(\underline{c},f(a}\delta_{b)\underline{d})}\\
&+u_4( f^{(6)}_{e(\underline{c},ab}\delta_{f\underline{d})}
+ f^{(6)}_{e(\underline{a},cd}\delta_{f\underline{b})})
+u_5 (  f^{(6)}_{e(b,a)(c}\delta_{d)f}+f^{(6)}_{e(d,c)(a}\delta_{b)f})\\
&+{v_1\over 2}(f^{(4)}_{gab(c}f^{(4)}_{d)efg}+f^{(4)}_{gcd(a}f^{(4)}_{b)efg})
+v_2f^{(4)}_{egab}f^{(4)}_{cdfg} 
+v_3f^{(4)}_{ega(c}f^{(4)}_{d)bfg}+(e\leftrightarrow f)\,.
\label{f6diffeqn:gen}
\fe
As we shall argue in the next section, the constant coefficients in \eqref{f4diffeqn:gen} and \eqref{f6diffeqn:gen} can be fixed using results from the type II/heterotic duality and heterotic string perturbation theory.

\section{An example of $f^{(4)}$ and $f^{(6)}$ from Type II/Heterotic duality}
\label{sec:example}

In Section~\ref{sec:diff}, we wrote down the differential constraints \eqref{f4diffeqn:gen} and \eqref{f6diffeqn:gen} on the 4- and 6-derivative four-point couplings $f^{(4)}$ and $f^{(6)}$ between the 21 tensor multiplets in 6$d$ $(2,0)$ supergravity, with undetermined {\it model-independent} constant coefficients.  To determine these coefficients, we can consider the specific example of four-point scattering amplitudes in type IIB string theory on K3.  In this section, we will relate the exact non-perturbative 4- and 6-derivative couplings in type IIB on K3 to a certain limit of the one- and two-loop amplitudes in the $T^5$ compactified heterotic string theory, via a chain of string dualities.  With explicit expressions for the heterotic amplitudes, we verify the differential constraints (see Appendix~\ref{app:check} for the detailed computations), and thereby determine the model-independent constant coefficients.

\subsection{Type II/Heterotic duality}
We consider type IIB string theory on $K3\times S^1_B$ with string coupling $g_B$, and circle radius $r_B$. The 6$d$ limit of interest corresponds to keeping $g_B \sim {\cal O}(1)$ while sending $r_B\rightarrow \infty$. We shall work in units with type II string tension $\A'=1$. By T-duality, we can equivalently look at type IIA string theory on $K3\times S^1_A$ with string coupling $g_A=g_B/r_B\sim r_A $ and circle size $r_A=1/r_B$. In terms of type IIA parameters, the 6$d$ limit corresponds to $g_A\sim r_A\rightarrow 0$. Now we use type IIA/heterotic duality to pass to heterotic string theory on $T^4\times S^1$ where the size of both $T^4$ and $S^1$ are of order $r_A$ in type II string units. Since the heterotic string is dual to a wrapped NS5 brane on $K3$, their tensions satisfy the relation $M_h\equiv 1/\ell_h\sim 1/ g_A  \sim 1/r_A $. The heterotic string coupling, on the other hand, can be fixed by matching the 6$d$ (or 5$d$) supergravity effective couplings
\ie
{1\over g_A^2}\sim {M_h^8 r_A^4\over g_h^2}
\fe
to be $g_h \sim 1/r_A\sim M_h$.\footnote{ One can also derive this using the equivalence between the IIA string and the wrapped heterotic NS5 brane on $T^4$.}  Hence in the limit where the circle of $K3\times S^1_B$ in the type IIB picture decompactifies $r_B\rightarrow\infty$, we have
\ie
g_h\sim M_h \rightarrow \infty
\fe
in the dual $T^5$ compactified heterotic string theory.

Under the duality, the 21 tensor multiplets of $(2,0)$ supergravity on $S^1_B$ are related to the 21 abelian vector multiplets of heterotic string on $T^5$. In particular the effective action of the tensor multiplets in the $(2,0)$ supergravity is captured by that of the vector multiplets in heterotic string. Let us now focus on the four-point amplitude of abelian vector multiplets in heterotic string on $T^5$. As we shall see in the next subsection, apart from the tree-level contribution at 2-derivative order due to supergraviton exchange, the four-point amplitudes at 4-derivative and 6-derivative orders receive contributions up to one-loop and two-loop, respectively. Furthermore we will argue that, in the limit of interest $
g_h\sim M_h \rightarrow \infty$,
  these couplings in the effective action are free from contributions at higher loop orders.

Relative to the tree-level contribution to the 2-derivative amplitude $f^{(2)}$, the 4-derivative $f^{(4)}$ coupling in the 6$d$ $(2,0)$ supergravity from type IIB on K3, which contains $H^4$ can be written as
\ie
{ f^{(4)} \over f^{(2)} } \sim \lim_{g_h \sim {1/\ell_h} \to \infty} \ell_h^2 \left( \beta_0 + \beta_1 g_h^2 + \beta_2 g_h^4 + \beta_3 g_h^6 + \dotsb \right),
\fe
where $\beta_n g_h^{2n}$ is the $n$-loop contribution.  By using type I/heterotic duality and confirmed by two-loop computation in \cite{Gross:1986mw, D'Hoker:1994yr, Stieberger:2002wk, D'Hoker:2005jc}, it has been argued that the $F^4$ coupling in heterotic string is does not receive contributions beyond one-loop, namely $\beta_n = 0$ for $n \geq 2$.  Therefore, we expect that $f^{(4)}$ is completely captured by the one-loop contribution $\beta_1 g_h^2 \ell_h^2$.\footnote{ Note that the tree-level contribution $\beta_0 \ell_h^2$ to the 4-derivative coupling vanishes in the limit of interest $g_h \sim 1/\ell_h \to \infty$. Similarly, the tree-level and one-loop contributions $\gamma_0 \ell_h^4$ and $\gamma_1 \ell_h^4 g_h^2$ to the 6-derivative coupling vanish in that limit.}  Indeed, we will see in Section \ref{sec:oneloop} that the heterotic one-loop amplitude \eqref{oneloopscalarfactor:sim} satisfies the differential constraint for the 4-derivative coupling \eqref{f4diffeqn:gen} in 6$d$ $(2,0)$ supergravity.

Likewise, the 6-derivative $f^{(6)}$ coupling which contains $D^2 H^4$ can be written as
\ie
{ f^{(6)} \over f^{(2)} } \sim \lim_{g_h \sim {1/\ell_h} \to \infty} \ell_h^4 \left( \gamma_0 + \gamma_1 g_h^2 + \gamma_2 g_h^4 + \gamma_3 g_h^6 + \dotsb \right).
\fe
We will see in Section \ref{sec:twoloop} that the two-loop contribution \eqref{f6} corresponding to the $\gamma_2 g_h^4 \ell_h^4$ term alone satisfies the differential constraint \eqref{f6diffeqn:gen} for the 6-derivative coupling in 6$d$ $(2,0)$ supergravity.  This strongly suggests that the $D^2F^4$ does not receive higher than two-loop contributions in the $T^5$ compactified heterotic string theory, though we are not aware of a clear argument.\footnote{ The consistency check with the differential constraints still allows for the possibility of shifting the 4- and 6-derivative coupling $f^{(4)}$ and $f^{(6)}$ by eigenfunctions of the covariant Hessian. However, we believe that $f^{(4)}$ and $f^{(6)}$ are given exactly by the low energy limit of the heterotic one- and two-loop contributions. The above  possibility can in principle be ruled out by studying the limit to 6$d$ $(2,0)$ SCFT, but we will not demonstrate it here.}


\subsection{Heterotic string amplitudes and the differential constraints}
In this subsection, we compute the four-point amplitude of scalars in the abelian vector multiplets in five dimensions, of heterotic string on $T^5$, or more precisely, heterotic string compactified on the Narain lattice $\Gamma_{21,5}$. As explained in the previous subsection, the coefficients of $F^4$ (or $(\partial\phi)^4$) and $D^2F^4$ (or $\partial^6\phi^4$) in five dimensions at genus one and genus two in heterotic string capture {\it exactly} the six dimensional effective couplings $f^{(4)}$ and $f^{(6)}$ of type IIB string theory on K3, expanded in the string coupling constant including the instanton corrections. These results can be extracted by slightly modifying the 10$d$ heterotic string amplitudes, computed by D'Hoker and Phong (see for instance (6.5) of \cite{D'Hoker:1994yr} and (1.22) of \cite{D'Hoker:2005jc}).
Furthermore, we fix the constant coefficients in the differential constraints \eqref{f4diffeqn:gen} and \eqref{f6diffeqn:gen} by explicitly varying the heterotic amplitudes with respect to the moduli fields.

\subsubsection{One-loop four-point amplitude}\label{sec:oneloop}

The scalar factor in the 4-gauge boson amplitude at one-loop takes the form\footnote{ The summation over spin structures has been effectively carried out already in this expression.}
\ie
{\cal A}_1 = \int_{{\cal F}_1} {d^2\tau\over \tau_2^2} {\tau_2^{5\over 2} \Theta_\Lambda(\tau,\bar\tau) \over  \Delta(\tau)} \prod_{i=1}^4 {{d^2 z_i }\over \tau_2} e^{{1\over 2}\sum_{i<j}  s_{ij} G(z_i,z_j)}  \left\langle \prod_{i=1}^4 j^{a_i}(z_i) \right\rangle_\tau.
\label{oneloopscalarfactor:raw}
\fe
Here $\Lambda$ denotes the even unimodular lattice $\Gamma_{21,5}$, $\Delta(\tau)=\eta(\tau)^{24}$ is the weight 12 cusp form of $SL(2,\mathbb Z)$, and $\Theta_\Lambda$ is the theta function of the lattice $\Lambda$ with modular weight $({21 \over 2}, {5 \over 2})$. $ j^a$ stand for the current operators associated to the 5$d$ Cartan gauge fields in the Narain lattice CFT and $G(z_i,z_j)$ is the scalar Green function  on the torus. The $z_i$ integrals are performed over the torus and the $\tau$ over ${\cal F}_1$ which is the fundamental domain of $SL(2,\mathbb Z)$ on $\mathbb H^2$. Note that the integrand has total modular weight $(2,2)$, and hence the integral is independent of the choice of the fundamental domain of $SL(2,\mathbb{Z})$.
To extract the $F^4$ coefficient, we can simply set $s_{ij}$ to zero in the above scalar factor \eqref{oneloopscalarfactor:raw}, and write
\ie
f^{(4)}_{a_1a_2a_3a_4} \equiv \left( \left.{\cal A}_1\right|_{F^4} \right)_{a_1a_2a_3a_4}&= \int_{{\cal F}_1}  {d^2\tau\over \tau_2^2} {\tau_2^{5\over 2} \Theta_\Lambda(\tau,\bar\tau) \over  \Delta(\tau)} \prod_{i=1}^4 {d^2 z_i\over \tau_2}   \left\langle \prod_{i=1}^4 j^{a_i}(z_i) \right\rangle_\tau
\\
&=  \left. {\partial^4\over  \partial y^{a_1}\cdots\partial y^{a_4}} \right|_{y=0} \int_{\cal F}  {d^2\tau\over \tau_2^2} {\tau_2^{5\over 2} \Theta_\Lambda(y|\tau,\bar\tau) \over  \Delta(\tau)},
\label{oneloopscalarfactor:sim}
\fe
where $y$ is a vector in $\mathbb{R}^{21}$, that lies in the positive subspace of the $\mathbb{R}^{21,5}$ in which the lattice $\Lambda$ is embedded. We have rewritten the four-point function of the currents as a fourth derivative on the theta function (see, for example, \cite{Polchinski:1998rr}). The theta function $\Theta_\Lambda$ is defined as
\ie
\Theta_\Lambda(y|\tau,\bar\tau) &= e^{{\pi \over 2\tau_2 }y\circ y } \sum_{\ell\in \Lambda} e^{\pi i \tau \ell_L^2 - \pi i \bar\tau \ell_R^2 + 2\pi i \ell\circ y}
\\
&=  e^{{\pi \over 2\tau_2 }y\circ y } \sum_{\ell\in \Lambda} e^{\pi i \tau \ell\circ\ell - 2 \pi \tau_2 \ell_R^2 + 2\pi i \ell\circ y}.
\fe
As we have argued previously, the $f^{(4)}$ coupling satisfies a differential constraint \eqref{f4diffeqn:gen} on $\cM$ due to supersymmetry. Given the explicit expression for $f^{(4)}$  \eqref{oneloopscalarfactor:sim} from the type II/heterotic duality, we can now  proceed to  fix the constant coefficients in \eqref{f4diffeqn:gen}. The above expression for the 4-derivative term $f^{(4)}$ has previously been determined in \cite{Kiritsis:2000zi}.

In the following we will explicitly parametrize the coset moduli space $\cM$ and write down the covariant Hessian $\nabla_{(e}\cdot\nabla_{f)}$. We will work in a trivialization of the $SO(21)$ vector bundle $\cV$. In particular, we shall identify the coordinates on the base manifold $\cM$ with variations of the embedding on the lattice $\Gamma_{21,5}$ in ${\mathbb R}^{21,5}$.

Let $e_I$ be a set of lattice basis vectors, $I=1,\cdots,26$, with the pairing $e_I\circ e_J = \Pi_{IJ}$ given by the even unimodular quadratic form of $\Gamma_{21,5}$. Let $P_+$ and $P_-$ be the linear projection operator onto the positive and negative subspace, $\mathbb{R}^{21}$ and $\mathbb{R}^5$, respectively, of $\mathbb{R}^{21,5}$. We can write $P_+ e_I$ as $(e_{Ia}^L)_{a=1,\cdots,21}$, and $P_-e_I$ as $(e_{Ii}^R)_{i=1,\cdots,5}$. 

We expand the lattice vectors into components $y=\widetilde y^I e_I$.  The left components of $y$ are then $y^a = \widetilde y^I e_{Ia}^L$. The requirement that $y$ lies in the positive subspace means that $\widetilde y^I e^R_{Ii}=0$. This constraint implies that, under a variation of the lattice embedding, while $y\circ y = y^a y^a$ stays invariant, $y$ itself must vary, and so does $\ell\circ y = \ell^I e^L_{Ia} y^a$.

From
\ie
\sum_{a=1}^{21} e^L_{Ia} e^L_{Ja} - \sum_{i=1}^5 e^R_{Ii}e^R_{Ji} = \Pi_{IJ},
\fe
we see that $(e^L_{Ia}, e^R_{Ii})$ is the inverse matrix of $\Pi^{IJ}(e^L_{Ja}, -e^R_{Ji})$.
Note that $e^R_{Ii}$ are specified by $e^L_{Ia}$ up to an $SO(5)$ rotation. $e^R_{Ii}$ by itself is subject to the constraint 
\ie\label{erc}
\Pi^{IJ} e^R_{Ii} e^R_{Jj} = -\delta_{ij}.
\fe
This constraint leaves $26\times 5 - 15$ independent components of $e^R_{Ii}$. The $SO(5)$ rotation of the negative subspace further removes 10 degrees of freedom from $e^R_{Ii}$, leaving the $21\times 5=105$ moduli of the lattice embedding which give rise to a parametrization of the scalar manifold $\cM$.\footnote{ Consider the symmetric matrix $M_{IJ}$ defined by
\ie
M_{IJ} = e^L_{Ia} e^L_{Ja} + e_{Ii}^R e_{Ji}^R = 2e^L_{Ia} e^L_{Ja}-\Pi_{IJ} = \Pi_{IJ} + 2 e_{Ii}^R e_{Ji}^R.
\fe
We have
\ie
M_{IK} \Pi^{KL} M_{LJ} = e^L_{Ia} \delta_{ab} e^L_{Jb} + e_{Ii}^R (-\delta_{ij}) e_{Jj}^R = \Pi_{IJ} .
\fe
The symmetric matrix $M$, subject to the constraint $M\Pi M=\Pi$, can be used to parameterize the coset $SO(21,5)/(SO(21)\times SO(5))$. }


Now consider variation of the lattice embedding,
\ie
e^L_{Ia}\to e^L_{Ia} + \delta e^L_{Ia},~~~ e^R_{Ii}\to e^R_{Ii} + \delta e^R_{Ii}.
\fe
subject to the constraints
\ie
& \Pi^{IJ} e^L_{I(a} \delta e^L_{J b)} = {\cal O}((\delta e)^2),~~~ \Pi^{IJ} \left( \delta e^L_{Ia} e^R_{J i} + e^L_{Ia} \delta e^R_{J i} \right) = {\cal O}((\delta e)^2),~~~ \Pi^{IJ} e^R_{I(i}\delta e^R_{Jj)} = {\cal O}((\delta e)^2),
\\
& e^L_{(Ia} \delta e^L_{J)a} - e^R_{(Ii} \delta e^R_{J)i} = {\cal O}((\delta e)^2).
\fe
Let $f(e^R_{Ii})$ be a scalar function on the moduli space $\cM$ of the embedding of $\Gamma_{21,5}$. We can expand
\ie\label{fo}
f(e^R_{Ii}+\delta e^R_{Ii}) = f(e) + f^{Ii}(e) \delta e^R_{Ii} + f^{IJij}(e) \delta e^R_{Ii}\delta e^R_{Jj}+{\cal O}((\delta e)^3).
\fe
$f^{Ii}$ and $f^{IJij}$ are subject to shift ambiguities 
\ie\label{shifta}
& f^{Ii} \to f^{Ii} + \Pi^{IJ} e^R_{Jj} g^{ij},
\\
& f^{IJij} \to f^{IJij} + {1\over 2} \Pi^{IJ} g^{ij} + \Pi^{IK} e^R_{Kk} h^{JKijk}+ \Pi^{JK} e^R_{Kk} h^{IKijk}
\fe
for arbitrary symmetric $g^{ij}$ and $h^{IJijk}$, due to the constraints on $\delta e^R_{Ii}$. We can fix these ambiguities by demanding 
\ie\label{ft}
& e^R_{Ij} f^{Ii}=0
,~~~~ e^R_{Ik} f^{IJij}=0. 
\fe
This can be achieved, for instance, by shifting $f^{Ii}$ with $\Pi^{IJ}e^R_{Jj} g^{ij}$, for some $g^{ij}$.
We can then define
\ie
\widetilde f_a{}^i = e^L_{I a} f^{Ii},~~~\widetilde f_{ab}{}^{ij} =  e^L_{Ia} e^L_{J b} f^{IJij} + {1\over 2} \delta_{ab} e^R_{I}{}^{(i} f^{Ij)}.
\fe
Note that these are invariant under the shift (\ref{shifta}) and hence give rise to well-defined differential operators on the moduli space $\cM$. 

This construction can be straightforwardly generalized  to non-scalar functions on $\cM$, and we can therefore write the covariant Hessian of $f^{(4)}_{abcd}$ \eqref{oneloopscalarfactor:sim} as
\ie
\nabla_{(e}\cdot \nabla_{f)} f^{(4)}_{abcd}= \sum_{i=1}^5 \tilde f^{(4)~~~~ii}_{abcd;ef}.
\fe
In Appendix \ref{app:f4}, we explicitly compute $\tilde f^{(4)~~~~ii}_{abcd;ef}$ and find
\ie
\sum_{i=1}^5 \tilde f^{(4)~~~~ii}_{abcd;ef}=-{3\over2} f^{(4)}_{abcd}\D_{ef}-2 f^{(4)}_{(e(abc}\D_{d)f)}+6f^{(4)}_{ef(ab}\D_{cd)}.\fe
which allows us to fix the constant coefficients in \eqref{f4diffeqn:gen} to be
\ie
U=-{3\over2},~~~~V=-2,~~~~W=6.
\fe

\subsubsection{Two-loop four-point amplitude}\label{sec:twoloop}
The scalar factor in the two-loop heterotic amplitude takes the form \cite{D'Hoker:2005jc}
\ie
{\cal A}_2 = \int_{{\cal F}_2} {\prod_{I\leq J} d^2\Omega_{IJ}\over (\det{\rm Im}\Omega)^{5\over 2} \Psi_{10}(\Omega)} \Theta_\Lambda(\Omega,\bar\Omega) \int_{\Sigma^4} e^{{1\over 2}\sum_{i<j}  s_{ij} G(z_i,z_j)} \left\langle \prod_{i=1}^4 j^{a_i}(z_i) \right\rangle_\Omega \overline{\cal Y}_S,
\fe
where ${\cal Y}_S$ is given by
\ie
& {\cal Y}_S = {1\over 3} (k_1-k_2)\cdot (k_3-k_4) \Delta(z_1,z_2) \Delta(z_3,z_4) + (2~{\rm permutations}),
\\
& \Delta(z,w) \equiv \epsilon^{IJ} \omega_I(z) \omega_J(w) .
\fe
$\omega_I(z)$ are a basis of holomorphic one-form on the genus two Riemann surface normalized such that
\ie
\oint_{\A_I}  \omega_J =\delta_{IJ},~~~~\oint_{\B_I} \omega_J = \Omega_{IJ},
\fe
where the cycles $\A_I$ and $\B_J$ have intersection numbers $(\A_I,\B_J)=\delta_{IJ}$, $(\A_I, \A_J)=(\B_I,\B_J)=0$.
$\Psi_{10}(\Omega)$ is the weight 10 Igusa cusp form of $Sp(4,\mathbb{Z})$ \cite{Moore:1986rh}. $\mathcal{F}_2$ is the moduli space of the genus two Riemann surface. $G(z_i, z_j)$ is the Green's function on the genus two Riemann surface. 
 Again, the contribution to $D^2F^4$ coefficient is simply extracted as
\ie
\left. {\cal A}_2\right|_{D^2F^4} &= {t-u\over 3} \int_{{\cal F}_2} {\prod_{I\leq J} d^2\Omega_{IJ}\over (\det{\rm Im}\Omega)^{5\over 2} \Psi_{10}(\Omega)} \Theta_\Lambda(\Omega,\bar\Omega) \int_{\Sigma^4}   \left\langle \prod_{i=1}^4 j^{a_i}(z_i) \right\rangle_\Omega\overline{\Delta(z_1,z_2) \Delta(z_3,z_4)} 
\\
&~~~ + (2~{\rm permutations}).
\fe
To proceed, we need to compute the four-point correlation function of
\ie
{\cal T}^a_I = \int {d^2 z} j^a(z) \overline{\omega_I(z)}= \oint_{\A_I} j^a(z) dz,
\fe
on the genus two Riemann surface $\Sigma$.
This allows us to express the correlators of ${\cal T}^a_I$ in terms of the theta function (see for instance \cite{Polchinski:1998rr}),
\ie
\Theta_\Lambda(\Omega, \bar\Omega) \left\langle \prod_{i=1}^n {\cal T}^{a_i}_{I_i} \right\rangle =
(\det\text{Im}\,\Omega)^2 \left. {\partial^n\over\partial y^{a_1}_{I_1}\cdots\partial y^{a_n}_{I_n}} \right|_{y=0} \Theta_\Lambda (y|\Omega,\bar\Omega),
\fe
where
\ie
\Theta_\Lambda (y|\Omega,\bar\Omega)& \equiv \sum_{\ell^1, \ell^2 \in\Lambda} e^{\pi i \Omega_{AB} \ell^A_L \cdot\ell^B_L - \pi i \bar\Omega_{AB}\ell^A_R\cdot\ell^B_R + 2\pi i \ell^A\circ y_A + {\pi\over 2}(({\rm Im}\Omega)^{-1})^{AB} y_A \cdot y_B }\\
&=\sum_{\ell^1, \ell^2 \in\Lambda} e^{i\pi  \Omega_{AB} \ell^A \circ\ell^B -2\pi \text{Im}\, \Omega_{AB}\ell^{A}_R\cdot \ell^{B}_R+ 2\pi i \ell^A\circ y_A + {\pi\over 2}(({\rm Im}\Omega)^{-1})^{AB} y_A \cdot y_B }.
\fe
Thus, we can simplify the result to
\ie
\left. {\cal A}_2\right|_{D^2 F^4} &= \left[ {t-u\over 3} \epsilon_{IJ} \epsilon_{KL}\left. {\partial^4\over \partial y_I^{a_1} \partial y_J^{a_2} \partial y_K^{a_3} \partial y_L^{a_4}} \right|_{y=0} + (2~{\rm perms}) \right] \int_{{\cal F}_2} {\prod_{I\leq J} d^2\Omega_{IJ}\over (\det{\rm Im}\Omega)^{1\over 2} \Psi_{10}(\Omega)}\Theta_\Lambda(y|\Omega,\bar\Omega).
\fe

Next, we would like to verify that the coefficient functions $f^{(6)}$ extracted from ${\cal A}_2$ obeys the differential constraint \eqref{f6diffeqn:gen}  
on the moduli space $SO(5,21)/(SO(5)\times SO(21))$, and also fix the precise coefficients thereof. 
 
In principle, it should be possible to show that $\Delta f^{(6)}$ is $f^{(6)}$ plus the integral of a total derivative on the moduli space ${\cal F}_2$ of the genus two Riemann surface $\Sigma$, which reduces to a boundary contribution where $\Sigma$ is pinched into two genus one surfaces. However, this calculation is somewhat messy so instead we will fix the coefficients of for $(f^{(4)})^2$ by comparison to similar differential constraints on the tensor branch of the 6$d$ $(2,0)$ SCFT.


We can write $\left. {\cal A}_2\right|_{D^2 F^4}$ as
\ie\label{f6def}
(\left. {\cal A}_2\right|_{D^2 F^4})_{a_1a_2a_3a_4}
=f^{(6)}_{a_1a_2,a_3a_4} s_{12} + f^{(6)}_{a_1a_3,a_2a_4} s_{13} + f^{(6)}_{a_1a_4,a_2a_3} s_{14}\,,
\fe
However, the definition \eqref{f6def} of $f^{(6)}_{a_1a_2,a_3a_4}$ is ambiguous because $s_{12}+s_{13}+s_{14}=0$. We fix this ambiguity by imposing
\ie\label{f6condition}
f^{(6)}_{a_1(a_2,a_3a_4)}=0.
\fe
Explicitly, $f^{(6)}_{a_1a_2a_3a_4}$ is given by,
\ie\label{f6}
f^{(6)}_{a_1a_2,a_3a_4}  &= {1\over 3} \left( \epsilon_{A_1A_3}\epsilon_{A_2A_4} + \epsilon_{A_1A_4}\epsilon_{A_2A_3} \right)\\
&\times
 \left.{\partial^4\over \partial y_{A_1}^{a_1} \partial y_{A_2}^{a_2} \partial y_{A_3}^{a_3} \partial y_{A_4}^{a_4}} \int_{{\cal F}_2} {\prod_{I\leq J} d^2\Omega_{IJ}\over (\det{\rm Im}\Omega)^{1\over 2} \Psi_{10}(\Omega)}\Theta_\Lambda(y|\Omega,\bar\Omega)\right|_{y=0} .
\fe
$f^{(6)}_{a_1a_2,a_3a_4}$ enjoys the symmetry
\ie\label{f6sym}
f^{(6)}_{a_1a_2,a_3a_4}=f^{(6)}_{a_2a_1,a_3a_4}=f^{(6)}_{a_1a_2,a_4a_3}=f^{(6)}_{a_3a_4,a_1a_2}\,.
\fe
The condition \eqref{f6condition} gives rise to constraints between the coefficients in \eqref{f6diffeqn:gen},
\ie\label{uup}
u_2 + {u_2'\over2}=0,~~u_3 + {u_3'\over2}=0.
\fe
We therefore end up with 5 coefficients $u_1,u_2,u_3,u_4,u_5$ to determine for the terms proportional to $f^{(6)}$ on the RHS of the differential equation \eqref{f6diffeqn:gen}. We determine the $u_i$'s by explicit computation of the covariant Hessian in Appendix \ref{app:f6} and find
\ie\label{ui}
u_1=-2,~~u_2=1,~~u_3=0,~~u_4=1,~~u_5=0.
\fe

On the other hand, to determine the 2 independent coefficients $v_1,v_2$ for the $(f^{(4)})^2$ terms in \eqref{f6diffeqn:gen}, we shall take advantage of the following differential constraint on the 6-derivative four-point term in the tensor branch effective action of $(2,0)$ SCFT~\cite{CDLY}
\ie
2\partial_{(e}\cdot \partial_{f)} F^{(6)}_{ab,cd}
=
&{w_1\over 2}(F^{(4)}_{gab(c}F^{(4)}_{d)efg}+F^{(4)}_{gcd(a}F^{(4)}_{b)efg})
+w_2 F^{(4)}_{egab}F^{(4)}_{cdfg} 
+
w_3 F^{(4)}_{ega(c}F^{(4)}_{d)bfg}+(e\leftrightarrow f)\,.
\label{f6diffeqn:20scft}
\fe
Here $F^{(6)}$ and $F^{(4)}$ are 6-derivative and 4-derivative four-point couplings of tensor multiplets in the $(2,0)$ SCFT, which are related to the supergravity couplings $f^{(6)}$ and $f^{(4)}$ by taking the large volume limit of $K3$ and zooming in on an ADE singularity that gives rise to the particular 6$d$ SCFT.\footnote{
The contribution due to supergraviton exchange on the RHS of \eqref{f6diffeqn:gen} is absent in  \eqref{f6diffeqn:20scft} due to this decoupling limit. A similar reduction of the genus one and two amplitudes in the type II string theory to supergravity amplitudes was considered in \cite{Tourkine:2013rda}.
} The 4- and 6-derivative terms on the tensor branch of the 6$d$ $(2,0)$ SCFT can be in turn computed by the one- and two-loop amplitudes in 5$d$ maximal SYM on its Coulomb branch as discussed in \cite{Cordova:2015vwa}. 
 Explicitly, we have (see Appendix \ref{app:f4sym} and \ref{app:f6sym} for details)
\ie
\left. {\cal A}_1\right|_{ F^4} \rightarrow  2^{10}\pi^{13\over 2} F^{(4)},
\quad
\left. {\cal A}_2\right|_{D^2 F^4} \rightarrow 2^{15}\pi^9 F^{(6)}.
\fe
In \cite{CDLY}, the coefficients in \eqref{f6diffeqn:20scft} are fixed to be\footnote{ The factor $2\pi$ in the numerator comes from the relative normalization between the lattice vectors and the 5$d$ scalars. In particular, the mass square of the $W$-boson is $m^2 = 2\pi \ell_R^2$.}
\ie
w_1=0,~~w_2=-w_3=-{2\pi\over 3\times 2^{11} \pi^4}.
\fe
Hence we fix the rest of the constants in \eqref{f6diffeqn:gen} to be
\ie
v_1=0,~~v_2=-v_3=-{1\over 3\times 2^{15} \pi^7}.
\fe

In summary, the 4- and 6-derivative couplings satisfy the following differential equations
\ie
\boxed{
\begin{aligned}
\nabla_{(e}\cdot \nabla_{f)} f^{(4)}_{abcd} &= -{3 \over 2} f^{(4)}_{abcd}\D_{ef} - 2 f^{(4)}_{(\underline{e}(abc}\D_{d)\underline{f})} + 6 f^{(4)}_{ef(ab}\D_{cd)},
\\
2\nabla_{(e}\cdot \nabla_{f)} f^{(6)}_{ab,cd}
&= -2 f^{(6)}_{ab,cd} \delta_{ef}
+ \left(   f^{(6)}_{ef ,ab}\delta_{cd}  +  f^{(6)}_{ef,cd}\delta_{ab} 
- 2  f^{(6)}_{ef ,(\underline{c}(a} \delta_{b) \underline{d})} \right) \\
&+ \left( f^{(6)}_{e(\underline{c},ab}\delta_{f\underline{d} }
+ f^{(6)}_{e(\underline{a},cd}\delta_{f\underline{b})} \right) - {1 \over 3 \times 2^{15} \pi^7} \left( f^{(4)}_{egab}f^{(4)}_{cdfg} 
- f^{(4)}_{ega(c}f^{(4)}_{d)bfg} \right) 
\\
&+(e\leftrightarrow f)\,.
\end{aligned}
}
\fe


\section{Implications of $f^{(4)}$ and $f^{(6)}$ for the K3 CFT}\label{sec:K3}

As alluded to in the introduction, since spacetime supersymmetry imposes differential constraints on the four-point string perturbative amplitudes which involve, in particular, integrated correlation functions of exactly marginal operators in the internal K3 CFT, we will be able to derive nontrivial consequences for the K3 CFT itself. As an illustration, we will see how the resulting moduli dependence of the $f^{(4)}$ and $f^{(6)}$ couplings at tree-level can pinpoint the singular points on the moduli space of the K3 CFT 
which the Zamolodchikov metric does not detect (since the moduli space is a symmetric space).

Below we will first explain how to extract the data relevant for K3 CFT from string tree-level amplitudes and general features thereof. We will then demonstrate their implications, in a particular slice of the K3 CFT moduli space where the K3 CFT is approximated by the supersymmetric nonlinear sigma model on $A_1$ ALE space.

In principle, we expect to arrive at the same set of constraints from the K3 CFT worldsheet Ward identities with spin fields associated to the Ramond sector ground states which appear in the spacetime supercharge. The same set of constraints is expected to hold for all $c= 6$ $(4,4)$ SCFTs.\footnote{
The condition $c=6$ is used in writing down the spacetime supercharges (in combination with the spacetime part of the worldsheet CFT) hence making connection to the spacetime supersymmetry constraints on the integrated CFT correlation functions.
}
We will leave this generalization to future work.

\subsection{Reduction to the K3 CFT moduli space}
The string theory amplitudes we have obtained in the previous section are exact results which can be regarded as sections of certain $SO(21)$ vector bundles over the full moduli space ${\cal O}(\Gamma_{21,5})\backslash SO(21,5)/SO(21)\times SO(5)$. Among the 105 moduli, one comes from the IIB dilaton, 24 come from RR fields, and the rest 80 NSNS scalars describe the moduli space of the K3 CFT, thus locally
\ie
{SO(21,5)\over SO(21)\times SO(5)}\approx {SO(20,4)\over SO(20)\times SO(4)}\times H^*(K3,{\mathbb R})\times {\mathbb R}^+.
\fe
Globally the K3 CFT has moduli space \cite{Seiberg:1988pf,Aspinwall:1994rg}
\ie
\cM_{K3}= {\cal O}(\Gamma_{20,4})\backslash SO(20,4)/SO(20)\times SO(4),
\fe 
parametrized by the scalars inside 20 of the 21 tensor multiplet, $\varphi^{\pm\pm}_i$ with $i=1,2\dots, 20$, from the 6$d$ perspective. From the worldsheet CFT point of view, $\varphi_i^{\pm\pm}$ are associated with the BPS superconformal primaries that are doublets of the two $SU(2)$ current algebras.

We will restrict the full string four-point amplitude obtained from the type II/heterotic duality to these 20 tensor multiplets, and expand in the limit of small $g_{\text {\tiny IIB}}$.\footnote{
Note that by doing so we break the $SO(5)$ R symmetry to $SU(2)\times SU(2)$.} In this limit, the theta function of the $\Gamma_{21,5}$ lattice can be approximated by the product of the theta function of the $\Gamma_{20,4}$ lattice and that of the $\Gamma_{1,1}$ lattice whose integral basis has the following embedding in ${\mathbb R}^{1,1}$,
\ie
u=(r_0,r_0),\quad v=({1\over 2r_0},-{1\over 2r_0}),
\fe
with $r_0\rightarrow \infty$ in the limit. Since at genus one $\Theta_{\Gamma_{1,1}}(\tau)\sim r_0$ and at genus two $\Theta_{\Gamma_{1,1}}(\Omega)\sim r_0^2$, we have in this limit 
\ie
\cA^{full}_{H^4}\sim r_0 \cA^{red}_{H^4}(\varphi_i),\quad
\cA^{full}_{D^2H^4}\sim r_0^2 \cA^{red}_{D^2 H^4}(\varphi_i).
\fe
Now on the other hand,  working with the canonically normalized fields (Einstein frame) which involves rescaling the string frame metric and $B$-fields by
\ie
G_{\m\n}\rightarrow M_6^{-2} G_{\m\n},~~B_{\m\n}\rightarrow M_6^{-2} B_{\m\n}
\fe
where $M_6=(V_{K3}/g_{\text{\tiny IIB}}^2 \ell_s^8)^{1/4}$ is the 6$d$ Planck scale, we know that the four-point coupling $f^{(4)}$ must scale as $M_6^2$ and $f^{(6)}$ as $M_6^4$. From this we conclude that $r_0\sim M_6^{2}$.


The differential constraint on $\cA^{full}$ in the perturbation expansion implies a similar constraint on $\cA^{red}$. Focusing on the  scalar component of the superamplitude, we have the following derivative expansion
\ie
&\cA^{red}_{H^4}( \varphi_i^{++} \varphi_j^{++} \varphi_k^{--} \varphi_\ell^{--})
\\
&=
 s^2 \left[ {\delta_{ij }\delta_{ k\ell}\over s}
+{\delta_{ik}\delta_{j\ell}\over t}+{\delta_{i\ell}\delta_{jk}\over u} 
+ A_{ijk\ell} + B_{ij,k\ell} s+B_{ik,j \ell} t+B_{i\ell,jk} u + {\cal O}(s^2)\right].
\label{treederexp}
\fe
where the first two terms come from the supergraviton exchange, while $A_{ijkl}$ and $B_{ij,kl}$ are obtained from the tree-level limit of the $f^{(4)}$ and $f^{(6)}$ couplings respectively. The coefficients $A_{ijk\ell}$ and $B_{ij,k\ell}$  for the $\mathcal{N}=4$ $A_{K-1}$ cigar CFT,
which is the $\mathbb{Z}_K$ orbifold of the supersymmetric $SU(2)_K/U(1)\times SL(2)_K/U(1)$ coset CFT, are studied in \cite{Chang:2014jta}.

On the other hand, $\cA_{red}$ can be evaluated directly from IIB  tree-level perturbation theory. The K3 CFT admits a small $(4,4)$ superconformal algebra, that contains left and right moving $SU(2)$ R-current algebra at level $k=1$ \cite{Eguchi:1987sm}. Focusing on the left moving part, the super-Virasoro primaries are labeled by its weight $h$ and $SU(2)$ spin  $\ell$. The BPS super-Virasoro primaries in the (NS,NS) sector consist of the identity operator ($h=\ell=\bar h=\bar\ell=0$), and 20 others labeled by ${\cal O}_i^{\pm\pm}$ with $h=\ell=\bar h=\bar\ell=1/2$ which correspond to the 20 $(1,1)$ harmonic forms in the K3 sigma model.\footnote{
Here $\mathcal{O}^{\pm\pm}_i$ are BPS superconformal primaries of the $\mathcal{N}=(4,4)$ superconformal algebra. With respect to an $\mathcal{N}=(2,2)$ superconformal subalgebra, ${\cal O}^{++}_i$ is a chiral primary  both on the left and the right, whereas ${\cal O}^{-+}_i$ is an anti-chiral primary on the left and a chiral primary on the right.
} 
The BPS primaries ${\cal O}_i^{\pm\pm}$ are the exactly marginal primaries of the K3 CFT, corresponding to the moduli fields $\varphi_i^{\pm\pm}$.
Under spectral flow, the identity operator is mapped to a unique $h=\bar h=1/4$, $\ell=\bar{\ell}=1/2$ ground state ${\cal O}^{\pm\pm}_0$ in the (R,R) sector, whereas the weight-$1/2$ BPS super-Virasoro primaries give rise to $h=\bar h=1/4$, $\ell=\bar \ell=0$ (R,R) sector ground states labeled by $\phi^{RR}_i$ \cite{Eguchi:1987sm}. The vertex operators for the 6$d$ massless fields all involve these 21 BPS super-Virasoro primaries and their spectral flowed partners.

 
 More explicitly, the vertex operators in the NSNS sector are
\ie\label{vo1}
&e^{-\phi -\bar\phi} \psi_\mu \bar \psi_\nu e^{ik \cdot X} \cdot 1,\\
& e^{-\phi -\bar\phi} e^{ik \cdot X} \cdot \mathcal{O}_i^{\pm\pm},~~~~i=1,\cdots,20,\\
\fe
Here $e^{ik\cdot X}$ comes from the $\mathbb{R}^{1,5}$ part of the worldsheet CFT. The associated 1-particle states transform under the $SU(2)\times SU(2)$ little group as
\ie\label{vertex1}
&{\bf (3,3) \oplus (3,1) \oplus (1,3) \oplus (1,1)}\, ,\\
&20\times 4\times {\bf (1,1)}\,.
\fe
The 80 scalars in the second line of \eqref{vo1} are denoted by $\varphi_i^{\pm\pm}$.

On the other hand, the vertex operators in the RR sector are
\ie
&e^{-\phi /2-\bar\phi/2} S_{\dot \A } \bar S_{\dot \B} e^{ik \cdot X} \cdot  \mathcal{O}_0^{\pm\pm},\\
& e^{-\phi /2-\bar\phi/2} S_\A \bar S_\B e^{ik \cdot X} \cdot \phi_i^{RR},~~~~i=1,\cdots,20.\\
\fe
The chiralities of the spin fields are dictated by the IIB GSO projection in RR sector, which depends on the $SU(2)$ R-charge of the vertex operator.\footnote{
 The IIB GSO projection in the RR sector is \cite{Giveon:1999px,Chang:2014jta} 
 \ie
 & F_L+2J^3_L-{1\over 2}\in 2{\mathbb{Z}},\quad F_R+2J^3_R-{1\over 2}\in 2{\mathbb{Z}}
 \fe
 where $F_{L,R}$ are the left and right worldsheet fermion numbers in $\mathbb{R}^{1,5}$, and $J^3_{L,R}$ denote the left and right $SU(2)$ Cartan R-charges of the internal K3 CFT.
 } The associated 1-particle states transform as
\ie\label{vertex2}
&4\times \Big(\,{\bf (1,3) \oplus (1,1)}\,\Big)\,,\\
&20\times\Big(\,  {\bf (3,1)\oplus(1,1)}\,\Big)\,,
\fe
under the $SU(2)\times SU(2)$ little group. \eqref{vertex1} and \eqref{vertex2} together give the 1-particles states in the (2,0) supergravity multiplet and the 21 tensor multiplets.  See Table \ref{table:3} for  summary.

\begin{table}
\centering
\begin{tabular}{|ccc|c|}
\hline
NSNS & & RR  & 6$d$ multiplet
\\\hline\hline
${\bf 1}_{h = \bar h = 0}$ & $\leftrightarrow$ & ${\bf 4}_{h = \bar h = {1 \over 4}}$  & supergravity + tensor
\\\hline
$20 \times {\bf 4}_{h = \bar h = {1 \over 2}}$ & $\leftrightarrow$ & $20 \times {\bf 1}_{h = \bar h = {1 \over 4}}$  & 20 tensors
\\\hline
\end{tabular}
\caption{The BPS primaries of the K3 CFT and the associated 6$d$ massless multiplets of type IIB string theory on $\bR^{1,5} \times K3$.
The $\bf 1$ and $\bf 4$ denote the trivial and $\ell = \bar \ell = {1 \over 2}$ representations of the worldsheet $SU(2)$ R-symmetry, and the arrows represent the spectral flow. }\label{table:3}
\end{table}

The four-scalar amplitude of $\varphi_i^{\pm\pm}$ in tree-level string theory is given by
\ie\label{ggz}
&\cA^{red}_{H^4}( \varphi_i^{++} \varphi_j^{++} \varphi_k^{--} \varphi_\ell^{--})
\\
&=\int {d^2 z \over 2\pi}\left\langle G_{-{1\over 2}}\overline G_{-{1\over 2}}{{\cal O}}^{++}_i e^{ik_1\cdot X}(z) \, G_{-{1\over 2}}\overline G_{-{1\over 2}} {{\cal O}}^{++}_j e^{ik_2\cdot X}(0) \,{{\cal O}}^{--}_ke^{ik_3\cdot X}(1)\, {{\cal O}}^{--}_\ell e^{ik_4\cdot X}(\infty) \right \rangle
\fe
where $G(z)$ is the ${\cal N} = 1$ super-Virasoro current, which acts on both ${\cal O}_i$ and $e^{ik\cdot X}$.\footnote{ The sigma model on $\bR^{1,5} \times K3$ has ${\cal N} = 2$ worldsheet supersymmetry.  The ${\cal N} = 1$ super-Virasoro current $G(z)$ is the sum of ${\cal N} = 2$ super-Virasoro currents $G^+(z) + G^-(z)$, and $G^\pm$ are each a combination of the ${\cal N} = 4$ super-Virasoro currents.  The $U(1)$ charge of the ${\cal N} = 2$ algebra coincides with the $J^3$ charge of the ${\cal N} = 4$.}  We have put two vertex operators in the $(-1,-1)$ picture and the other two in the $(0,0)$ picture to add up to the total picture number $(-2,-2)$ for the tree-level string scattering amplitude. The correlator of the superconformal ghosts have already been taken into account in the above.

By deforming the contour of  $G^-_{-{1\over 2}}=\oint {dw\over 2\pi i}G^-(w)$, it is easy to see that the following correlation function vanishes identically
\ie
& \left\langle G_{-{1\over 2}}\overline G_{-{1\over 2}}{{\cal O}}^{++}_i(z) \, G_{-{1\over 2}}\overline G_{-{1\over 2}} {{\cal O}}^{++}_j(0) \,{{\cal O}}^{--}_k(1)\, {{\cal O}}^{--}_\ell(\infty) \right \rangle
\\
&= \left\langle G^-_{-{1\over 2}}\overline G^-_{-{1\over 2}}{{\cal O}}^{++}_i(z) \, G^-_{-{1\over 2}}\overline G^-_{-{1\over 2}} {{\cal O}}^{++}_j(0) \,{{\cal O}}^{--}_k(1)\, {{\cal O}}^{--}_\ell(\infty) \right \rangle.
\fe
Therefore in \eqref{ggz} we can take $G_{-{1\over 2}},\overline G_{-{1\over 2}}$ to act on $e^{ik\cdot X}$ only, which gives
\ie
\cA^{red}_{H^4}( \varphi_i^{++} \varphi_j^{++} \varphi_k^{--} \varphi_\ell^{--})= s^2 \int {d^2 z\over 2\pi}\, |z|^{-{ s  }-2} |1-z|^{-{  t }}\left\langle {{\cal O}}^{++}_i(z) {{\cal O}}^{++}_j (0) {{\cal O}}^{--}_k(1) {{\cal O}}^{--}_\ell (\infty) \right \rangle.
\fe
Thus, comparing with \eqref{treederexp}, we obtain the relation 
\ie\label{4ptf&amp}
&  \int {d^2 z\over 2\pi}\, |z|^{-{  s }-2} |1-z|^{-{  t }}\left\langle {\cal O}^{++}_i(z) {\cal O}^{++}_j (0) {\cal O}^{--}_k(1) {\cal O}^{--}_\ell (\infty) \right \rangle
\\
&=   {\delta_{ij }\delta_{ k\ell}\over s}
+{\delta_{ik}\delta_{j\ell}\over t}+{\delta_{i\ell}\delta_{jk}\over u} 
+ A_{ijk\ell} + B_{ij,k\ell} s+B_{ik,j \ell} t+B_{i\ell,jk} u + {\cal O}(s^2,t^2,u^2) .
\fe
From the CFT perspective, 
the polar terms in $t$ and $u$ are simply due to the appearance of the identity operator in the OPE of ${\cal O}^{++}$ with ${\cal O}^{--}$, while  $A_{ijk\ell}$ and $B_{ijk\ell}$ capture information about all intermediate primaries in the conformal block decomposition of the four-point function of the marginal operators. It is then natural to expect this relation to hold for exactly marginal operators in any $c=6$ $(4,4)$ SCFT. Furthermore, we expect $A_{ijk\ell}$ and $B_{ijk\ell}$ to obey the same kind of differential equations as $f^{(4)}$ and $f^{(6)}$, for any $c=6$ $(4,4)$ SCFT.

Using the relation between the correlation function of ${\cal O}_i^{\pm\pm}$ and their spectral flowed partners $\phi_{i}^{RR}$, 
\ie
\left\langle {\cal O}^{++}_i(z) {\cal O}^{++}_j (0) {\cal O}^{--}_k(1) {\cal O}^{--}_\ell (\infty) \right \rangle
=
{|z|\over |1-z|}\left\langle \phi^{RR}_i(z) \phi^{RR}_j (0) \phi^{RR}_k(1) \phi^{RR}_\ell (\infty) \right \rangle,
\fe
we can put \eqref{4ptf&amp} into an equivalent form, where the crossing symmetries are manifest in all channels,
\ie\label{4ptf&amp}
&  \int {d^2 z\over 2\pi}\, |z|^{-{  s }-1} |1-z|^{-{  t }-1}\left\langle \phi^{RR}_i(z) \phi^{RR}_j (0) \phi^{RR}_k(1) \phi^{RR}_\ell (\infty) \right \rangle
\\
&=  {\delta_{ij }\delta_{ k\ell}\over s}
+{\delta_{ik}\delta_{j\ell}\over t}+{\delta_{i\ell}\delta_{jk}\over u} 
+ A_{ijk\ell} + B_{ij,k\ell} s+B_{ik,j \ell} t+B_{i\ell,jk} u + {\cal O}(s^2,t^2,u^2) .
\fe

\subsection{$A_1$ ALE Limit}
To illustrate the power of the relation \eqref{4ptf&amp}, we consider the $A_1$ ALE limit where we zoom in on and resolve an $A_1$ singularity. In other words, we focus on a slice near the boundary of the full moduli space ${\cal M}_{K3}$, where the K3 CFT is reduced to a sigma model on $A_1$ ALE space, which is related to the sigma model on ${\mathbb C}^2/{\mathbb Z}_2$ by exactly marginal deformations \cite{Dixon:1985jw, Cecotti:1990kz, Aspinwall:1994rg}.

The slice of interest is parametrized by the normalizable exactly marginal deformations of the orbifold CFT $\mathbb{C}^2/\mathbb{Z}_2$, which is simply the moduli space of the $A_1$ SCFT\footnote{
In \cite{Dijkgraaf:1998gf}, the moduli space of the non-linear sigma model on a general hyperk\"ahler manifold is discussed. For the $A_1$ ALE space sigma model, the moduli space metric is flat because we have scaled the Zamolodchikov metric by an infinite volume factor of the target space.}
\ie
\cM_{A_1}={\mathbb{R}^3\times S^1\over \mathbb{Z}_2},
\fe
where $\mathbb{R}^3$ corresponds to the K\"ahler and complex structure deformations associated with the exceptional divisor of the $\mathbb{C}^2/\mathbb{Z}_2$, and the $S^1$ is parameterized by the integral of the $B$ field on the exceptional divisor.
This $\mathbb{Z}_2$ can be understood from the fact that the $SO(3)$ rotation of the asymptotic geometry of the circle fibration of the Eguchi-Hanson geometry that exchanges the two points of degenerate fiber effectively also flips the orientation of the $\mathbb{P}^1$ hence reflects the $B$-field flux.
The two orbifold singularities on the moduli space corresponds to the free orbifold point and the singular CFT point where a linear dilaton throat develops. The distinction between these two points on the moduli space is not detected by the Zamolodchikov metric, but should be detected by $f^{(4)}$ restricted to the single tensor multiplet corresponding to this exceptional divisor (or rather $A_{1111}$).\footnote{
Note that $f^{(6)}$ vanishes in this case because there is only one tensor multiplet involved.}

Since the overall volume of the CFT target space is infinite, $A_{1111}$ is a harmonic function on the moduli space.\footnote{
The contribution from supergraviton exchange on the RHS of \eqref{f4diffeqn:gen} is suppressed in this limit.
} Near the singular CFT point, $A_{1111}$ goes like $1/| \vec \varphi|^2$, where $ \vec \varphi $ is a local Euclidean coordinate on the moduli space, as in the case of the $A_1$ DSLST at tree-level (either $(2,0)$ or $(1,1)$) \cite{Aharony:2003vk, Chang:2014jta}. At the free orbifold point, on the other hand, the four-point function of marginal operators are perfectly non-singular, and $A_{1111}$ should be finite. This together with the harmonicity and R-symmetry determines $A_{1111}$ to be (up to an overall coefficient)
\ie
A_{1111} = \sum_{n=-\infty}^\infty {1\over \sum_{i=1}^3\varphi_i^2 + (\varphi_4 - 2\pi n R)^2},
\label{A1f4}
\fe
where $R$ is the radius of the $S^1$ of the moduli space.\footnote{
The $S^1$ parameterized by the $B$-field flux through the exceptional divisor $\mathbb P^1$ is of constant size along the $\bR^3$. This is because the marginal primary operator associated with the normalizable harmonic 2-form on the ALE space with unit integral on the $\mathbb P^1$ also has a normalized two-point function.
} It is easy to identify from \eqref{A1f4} that, $\vec\varphi=(0,0,0,0)$ is the singular CFT point, and $\vec\varphi=(0,0,0,\pi R)$ is the free orbifold point, since $A_{1111}$ is non-singular at the latter point, and the $\mathbb{Z}_2$ symmetry is clearly preserved.

Let us define $r^2=\sum_{i=1}^3 \varphi_i^2$, $\varphi_4 = \pi R + y$. Then near the free orbifold point, $r,y$ are small, we have
\ie
A_{1111} = {1\over 4R^2} + {3y^2-r^2\over 48 R^4} + {\cal O}(r^4,y^4,r^2y^2).
\fe
One should be able to confirm this using conformal perturbation theory \cite{Lunin:2000yv}.

In the large $\varphi$ regime, where the CFT is described by a nonlinear sigma model on $T^*\mathbb{CP}^1$, performing Poisson summation on \eqref{A1f4}, we can write $A_{1111}$ as the expansion
\ie
\label{instanton}
A_{1111} = {1\over 2 R r} \left[ 1 + \sum_{n=1}^\infty (-)^n e^{-{n(r+iy)\over R}} + \sum_{n=1}^\infty (-)^n e^{-{n(r-iy)\over R}}  \right].
\fe
Since $r$ scales like the area of the $\mathbb{CP}^1$, the leading $1/r$ contribution should come from one-loop order in $\A'$ perturbation theory. The $e^{-nr/R}$ corrections, on the other hand, are expected to come from worldsheet instanton effects. Moreover, the phase $e^{\pm i n y/R}$ indicates that there are contributions from both holomorphic and anti-holomorphic worldsheet instantons. In other words, our exact result based on supersymmetry constraints gives the striking prediction that in $\A'$ perturbation theory, $A_{1111}$ which is related to the four-point function of exactly marginal operators of the $A_1$ SCFT,
receives only one-loop plus worldsheet instanton contributions.

It would be interesting to understand if a similar worldsheet instanton expansion applies for the K3 CFT at finite overall volume, and its relation to the ${\cal N}=4$ topological string \cite{Berkovits:1994vy, Antoniadis:2006mr, Antoniadis:2007cw, Gaberdiel:2011qu}. In particular, the $\mathcal{N}=4$ topological string amplitudes are written as integrals over the fundamental domain $\mathcal{F}_1$ and also satisfy certain differential equations on the moduli space \cite{Gaberdiel:2011qu}.


\section{Discussions}

The main result of this paper is the exact non-perturbative coupling of tensor multiplets at 4 and 6-derivative orders in type IIB string theory compactified on K3, $f^{(4)}_{abcd}(\phi)$ and $f^{(6)}_{ab,cd}(\phi)$, and the differential equations they obey on the 105-dimensional moduli space. In the weak coupling limit (tree-level string theory), as described in section 5, they reduce to (up to a factor involving the IIB string coupling) the functions $A_{ijk\ell}(\varphi)$ and $B_{ij,k\ell}(\varphi)$ on the 80-dimensional moduli space of the K3 CFT. $A_{ijk\ell}$ and $B_{ij,k\ell}$ are integrated four-point functions of ${1\over 2}$-BPS operators in the K3 CFT on the sphere. Unlike the Zamolodchikov metric or its curvature \cite{Kutasov:1988xb}, $A_{ijk\ell}$ and $B_{ij,k\ell}$ do {\it not} receive contribution from contact terms, and depend nontrivially on the moduli. In particular, these functions diverge at the points in the moduli space where the CFT develops a continuous spectrum (corresponding to ADE type singularities on the K3 surface, with no $B$-field through the exceptional divisors \cite{Aspinwall:1994rg}). This allows us to pinpoint the location on the moduli space using CFT data alone (as opposed to, say, BPS spectrum of string theory), and makes it possible to study the K3 CFT through the superconformal bootstrap \cite{Rattazzi:2010gj, Poland:2010wg, ElShowk:2012ht, Beem:2015aoa} (e.g. constraining the non-BPS spectrum of the CFT) at any given point on its moduli space. This is currently under investigation \cite{LSSWY}.

In the full type IIB string theory on K3, at the ADE points on the moduli space, there are new strongly interacting massless degrees of freedom, characterized by the 6$d$ $(2,0)$ superconformal theory at low energies. Near these points, the components of $f^{(4)}_{abcd}(\phi)$ and $f^{(6)}_{ab,cd}(\phi)$ associated with the moduli that resolve the singularities are precisely the $H^4$ and $D^2H^4$ couplings on the tensor branch of the $(2,0)$ SCFT, studied in \cite{Maxfield:2012aw, Cordova:2015vwa}. Note that this is different from the ALE space limit discussed in section 5.2, which was restricted to the weak string coupling regime.

As pointed out in Section \ref{sec:supervertex}, there are F-term supervertices involving the supergraviton in 6$d$ $(2,0)$ supergravity theories as well, including one that corresponds to a coupling of the schematic form $f_R(\phi)R^4+\cdots$. It appears that a six-point supervertex involving 4 supergravitons and 2 tensor multiplets in the $SO(5)_R$ singlet does not exist at this derivative order (namely 8), and so by the same reasoning as Section \ref{sec:diff}, we expect that $f_R(\phi)$ obeys a second order differential equation with respect to the moduli, whose form is determined by the factorization structure of the six-point superamplitude of 4 supergravitons and 2 tensor multiplets. One complication here is the potential mixing of the coefficients of $R^4$, $D^2(R^2H^2)$, and $D^4H^4$, in the differential constraining equations. In particular, $D^4H^4$ is a D-term, and by itself is not subject to such constraining equations. We leave a detailed analysis of the supersymmetry constraints on the higher derivative supergraviton couplings in $(2,0)$ supergravity to future work.

One can similarly classify the supervertices in the 6$d$ (1,1) supergravity theory and derive differential constraints for the higher derivative couplings. In this case however, the string coupling lies in the 6$d$ supergraviton multiplet rather than the vector multiplets, and its dependence is not controlled by the same type of differential equations considered in this paper.

Finally, one may wonder whether our exact results for integrated correlators in the K3 CFT can be extended to 2$d$ $(4,4)$ SCFTs with $c=6k$ for $k>1$, such as the D1-D5 CFT \cite{Seiberg:1999xz, Larsen:1999uk}. While this is conceivable, the arguments used in this paper are based on the spacetime supersymmetry of the string theory and cannot be applied directly to the $k>1$ case. In the CFT language, our constraints can be recast as Ward identities involving insertions of spin fields, and we have implicitly used the property that the spin fields of the $c=6$ $(4,4)$ SCFT transform in a doublet of the $SU(2)_R$ symmetry. It would be interesting to understand whether there are analogous Ward identities in the $c=6k$ $(4,4)$ SCFTs, where the spin fields carry $SU(2)_R$ spin $j={k\over 2}$.\footnote{
In the case of the $c=12$ $(4,4)$ SCFT, say described by the nonlinear sigma model on a hyperK\"ahler 4-fold, one may compactify type IIB string theory to 2$d$, which generally leads to a $(6,0)$ supergravity theory in two dimension \cite{Gates:2000fj, Sriharsha:2006in}, and examine the 4-derivative F-term coupling of moduli fields in this theory. However, we are not able to derive differential constraining equations on these couplings based on soft limits of superamplitudes, due to the existence of local supervertices for the relevant six-point couplings at the same derivative order, in contrast to the 6$d$ $(2,0)$ supergravity theory. }

\section*{Acknowledgments}

We would like to thank Clay C\'ordova, Thomas Dumitrescu, Hirosi Ooguri, David Simmons-Duffin, Cumrun Vafa for discussions.  We would like to thank the Quantum Gravity Foundations program at Kavli Institute for Theoretical Physics, the workshop ``From Scattering Amplitudes to the Conformal Bootstrap" at Aspen Center for Physics, and the Simons Summer Workshop in Mathematics and Physics 2015 for hospitality during the course of this work. SHS is supported by a Kao Fellowship at Harvard University. YW is supported in part by the U.S. Department of Energy under grant Contract Number DE-SC00012567. XY is supported by a Sloan Fellowship and a Simons Investigator Award from the Simons Foundation. 

\appendix 

\section{Explicit check of the differential constraints}
\label{app:check}

\subsection{Four-derivative coupling $f^{(4)}$}\label{app:f4}

In this Appendix we will explicitly show that the 4-derivative term coefficient $f^{(4)}_{abcd}$ between the 21 tensor multiplets satisfies the following differential equation and determine the coefficients $U,V,W$,
\ie\label{UVW}
\nabla_e \cdot \nabla_f f^{(4)}_{abcd} = U f^{(4)}_{abcd} \delta_{ef} +Vf^{(4)}_{( e(abc} \delta_{d)f)}
+W f^{(4)}_{ef (ab} \delta_{cd)}\,.
\fe

Let us first decompose the 4-derivative coefficient $f^{(4)}_{abcd}$ into the 
\ie\label{ABC}
f^{(4)}_{abcd} = A_{abcd}+ \delta_{(ab} B_{cd)} +\delta_{(ab} \delta_{cd)} C
\fe
where $A_{abcd}$ and $B_{cd}$ are symmetric and traceless. The covariant Hessian $\nabla_{(a}\cdot \nabla_{b)}$ of these tensors can be expressed, through a set of relations similar to (\ref{fo}) and (\ref{ft}), in the form
\ie
\sum_{i=1}^5\widetilde A_{abcd,efii}, ~~~\sum_{i=1}^5 \widetilde B_{ab,cdii},~~~ \sum_{i=1}^5\widetilde C_{abii}.
\fe 
The differential constraints \eqref{UVW} can be expressed as
\ie\label{ABCdiff}
& \sum_{e,i} \widetilde A_{abcd,eeii} = a  A_{abcd},
\\
& \sum_{c,i} \widetilde B_{ab,ccii} = b  B_{ab},
\\
& \sum_{a,i} \widetilde C_{aaii} = c  C,
\\
& \sum_i \widetilde A_{abcd,efii} - {1\over 21}\delta^{ef} \sum_{g,i} \widetilde A_{abcd,ggii} = u \delta_{(\underline{e}(a} A_{\underline{f})cde)} + v \delta_{e(\underline{a}} \delta_{f\underline{b}} B_{\underline{cd})} - {\rm traces},
\\
& \sum_i \widetilde B_{ab,cdii} - {1\over 21}\delta_{cd} \sum_{e,i} \widetilde B_{ab,eeii} = x A_{abcd} + y \delta_{(\underline{c}(a} B_{\underline{d})b)} + z \delta_{c({a}} \delta_{{b})d} C - {\rm traces},
\\
& \sum_i \widetilde C_{abii} - {1\over 21}\delta_{ab} \sum_{c,i} \widetilde C_{ccii} = w B_{ab}.
\fe
We will relate the coefficients $a,b,c,u,v,x,y,z,w$ to $U,V,W$ later.

To start with, let us determine the constant in the differential equation for the scalar function $C$. From \eqref{oneloopscalarfactor:sim} and \eqref{ABC}, we first write $C$ as
\ie
C &= 4! \int_{\cal F} {d^2\tau\,\tau_2^{1\over 2}\over\Delta(\tau)} \sum_{\ell\in\Lambda} q^{\ell\circ\ell\over 2} e^{-2\pi \tau_2 \ell^I e^R_{Ii}e^R_{Ji} \ell^J} \left[ {\pi^2\over 8\tau_2^2} - {\pi^3\over  21\tau_2} \ell^I \ell^J e^L_{Ia}e^L_{Ja} + {2\pi^4\over 3\cdot 161} (\ell^I \ell^J e^L_{Ia}e^L_{Ja})^2 \right]
\\
&= 24\int_{\cal F} {d^2\tau\,\tau_2^{1\over 2}\over\Delta(\tau)} \sum_{\ell\in\Lambda} q^{\ell\circ\ell\over 2}  \left[ {\pi^2\over 8\tau_2^2} - {\pi^3\over  21\tau_2} \left(\ell\circ\ell + {i\over \pi}  \partial_{\overline\tau}\right) + {2\pi^4\over 3\cdot 161} \left(\ell\circ\ell + {i\over \pi}  \partial_{\overline\tau} \right)^2 \right] e^{-2\pi \tau_2 \ell^I e^R_{Ii}e^R_{Ji} \ell^J},
\fe
where we have used $\ell^I\ell^Je^L_{Ia} e^L_{Ja} = \ell\circ \ell +\ell^I\ell^Je^R_{Ii} e^R_{Ji}$.  
After integration by part, we have
\ie
C
&= {16\pi^4\over 161}\int_{\cal F} {d^2\tau\,\tau_2^{1\over 2}\over\Delta(\tau)} \sum_{\ell\in\Lambda} q^{\ell\circ\ell\over 2}  
\Bigg[
(\ell\circ\ell)^2-{11\over \pi \tau_2}\ell\circ \ell+{33\over \pi^2 \tau_2^2}
\Bigg]e^{-2\pi \tau_2 \ell^I e^R_{Ii}e^R_{Ji} \ell^J}.
\fe
Under the variation $e^R\to e^R + \delta e^R$, the first and second order variations of $C$ are given by
\ie
&C^{Ii} \delta e^R_{Ii} + C^{IJij} \delta e^R_{Ii} \delta e^R_{Jj} 
\\
&= 24\int_{\cal F} {d^2\tau\,\tau_2^{1\over 2}\over\Delta(\tau)} \sum_{\ell\in\Lambda} q^{\ell\circ\ell\over 2}  \left[ {\pi^2\over 8\tau_2^2} - {\pi^3\over  21\tau_2} \left(\ell\circ\ell - {1\over 4\pi\tau_2} \right) \right.\\
&\left.~~~~+ {2\pi^4\over 3\cdot 161} \left( (\ell\circ\ell)^2 + {\ell\circ\ell\over 2\pi\tau_2} - {1\over 16\pi^2\tau_2^2} \right) \right] e^{-2\pi \tau_2 \ell^I e^R_{Ii}e^R_{Ji} \ell^J} 
\\
&~~~\times \left\{  - 4\pi \tau_2 \ell^I \ell^J e^R_{Ii} \delta e^R_{Ji} + \Big( -2\pi\tau_2 \ell^I  \ell^J \delta^{ij} + 8\pi^2\tau_2^2 \ell^I \ell^J \ell^K \ell^L e^R_{Ki} e^R_{Lj} \Big) \delta e^R_{Ii} \delta e^R_{Jj} \right\} .
\fe
We can thus determine
\ie
&\widetilde C_{abij} = e^L_{Ia} e^L_{Jb} C^{IJij} + {1\over 2} \delta_{ab} e^R_I{}^{(i} C^{Ij)} 
\\
&~~~~~ = 24\int_{\cal F} {d^2\tau\,\tau_2^{1\over 2}\over\Delta(\tau)} \sum_{\ell\in\Lambda} q^{\ell\circ\ell\over 2}  \left[ {\pi^2\over 8\tau_2^2} - {\pi^3\over  21\tau_2} \left(\ell\circ\ell - {1\over 4\pi\tau_2} \right) + {2\pi^4\over 3\cdot 161} \left( (\ell\circ\ell)^2 + {\ell\circ\ell\over 2\pi\tau_2} - {1\over 16\pi^2\tau_2^2} \right) \right] 
\\
&~~~~~~~~~\times e^{-2\pi \tau_2 \ell^I e^R_{Ii}e^R_{Ji} \ell^J} \left\{ -2 \pi \tau_2 \delta_{ab} \ell^I \ell^J e^R_{Ii} e^R_{Jj} + \Big( -2\pi\tau_2 \ell^I  \ell^J \delta_{ij} + 8\pi^2\tau_2^2 \ell^I \ell^J \ell^K \ell^L e^R_{Ki} e^R_{Lj} \Big) e^L_{Ia} e^L_{Jb} \right\} .
\fe
We can now compute the Laplacian of $C$,
\ie
&\sum_{a=1}^{21}\nabla_{a} \cdot \nabla_a  C= \sum_{a,i} \widetilde C_{aa ii} \\
&=   24\int_{\cal F} {d^2\tau\,\tau_2^{1\over 2}\over\Delta(\tau)} \sum_{\ell\in\Lambda} q^{\ell\circ\ell\over 2}  \left[ {\pi^2\over 8\tau_2^2} - {\pi^3\over  21\tau_2} \left(\ell\circ\ell - {1\over 4\pi\tau_2} \right) + {2\pi^4\over 3\cdot 161} \left( (\ell\circ\ell)^2 + {\ell\circ\ell\over 2\pi\tau_2} - {1\over 16\pi^2\tau_2^2} \right) \right] 
\\
&\times \left\{  {26}\tau_2\partial_{\tau_2} + \ell\circ \ell \Big( -10\pi\tau_2 -4 \pi\tau_2^2\partial_{\tau_2}\Big) + 2\tau_2^2 \partial_{\tau_2}^2 \right\} e^{-2\pi \tau_2 \ell^I e^R_{Ii}e^R_{Ji} \ell^J}  .
\fe
After replacing $\partial_{\tau_2}$ by $-2i\partial_{\overline\tau}$, and integration by parts, 
we find
\ie\label{Cdiff}
\sum_{a,i} \widetilde C_{aaii} = {25\over 2} C.
\fe
This fixes the constant $c$ in \eqref{ABCdiff} to be $c=25/2$.

Similarly we can write
\ie
	 A_{abcd} &={(2\pi i)^4 } \int_{\cal F} {d^2\tau\,\tau_2^{1\over 2}\over\Delta(\tau)} \sum_{\ell\in\Lambda} q^{\ell\circ\ell\over 2} e^{-2\pi \tau_2 \ell^I e^R_{Ii}e^R_{Ji} \ell^J}
\Bigg[
 \ell^I \ell^J\ell^K \ell^L e^L_{I\hat a}e^L_{J\hat b}e^L_{K\hat c}e^L_{L\hat d} 
\Bigg],\\
B_{ab}
&={96\pi^4\over 25}\int_{\cal F} {d^2\tau\,\tau_2^{1\over 2}\over\Delta(\tau)} \sum_{\ell\in\Lambda} q^{\ell\circ\ell\over 2} e^{-2\pi \tau_2 \ell^I e^R_{Ii}e^R_{Ji} \ell^J} \Bigg[
\ell\circ\ell- {6\over\pi\tau_2} 
	 \Bigg]\ell^I \ell^J e^L_{I\hat a}e^L_{J\hat b},
\fe
where the hatted indices are taken to be symmetric traceless combinations.

The covariant Hessians of $A_{abcd}$, $B_{ab}$, and $C$ can be computed straightforwardly to be
\ie
&\nabla_{(e}\cdot \nabla_{f)} A_{abcd}=\widetilde A_{abcd,efii}\\
  =&{16\pi^4}\int_{\cal F} {d^2\tau\,\tau_2^{1\over 2}\over\Delta(\tau)} \sum_{\ell\in\Lambda} q^{\ell\circ\ell\over 2} e^{-2\pi \tau_2 \ell^I e^R_{Ii}e^R_{Ji} \ell^J}
  \Bigg[
    - 2  \ell^I \ell^J\ell^M \ell^N e^L_{I\hat a}e^L_{J\hat b}e^L_{M\hat c}   e^L_{N (e}\D_{f)\hat d}
          +
    {3\over 2\pi\tau_2}\ell^I \ell^J  e^L_{I\hat a}e^L_{J\hat b}  \D_{\hat c e}\D_{\hat d f}
          \Bigg]
   \\
& -{24\pi^4}\D_{ef}\int_{\cal F} {d^2\tau\,\tau_2^{1\over 2}\over\Delta(\tau)} \sum_{\ell\in\Lambda} q^{\ell\circ\ell\over 2} e^{-2\pi \tau_2 \ell^I e^R_{Ii}e^R_{Ji} \ell^J}
  \ell^I \ell^J\ell^M \ell^N e^L_{I\hat a}e^L_{J\hat b}e^L_{M\hat c}e^L_{N\hat d},\\
&\nabla_{(c} \cdot \nabla_{d)} B_{ab} = \tilde B_{ab,cdii}\\
 =&{96\pi^4\over 25}\int_{\cal F} {d^2\tau\,\tau_2^{1\over 2}\over\Delta(\tau)} \sum_{\ell\in\Lambda} q^{\ell\circ\ell\over 2} e^{-2\pi \tau_2 \ell^I e^R_{Ii}e^R_{Ji} \ell^J}
\\
&\times \Bigg[ \ell^I \ell^J e^L_{I\hat a}e^L_{J\hat b} \Bigg(
24\ell^M \ell^N  e^L_{Mc}e^L_{Nd}
+\D_{cd}\left(
{3 \over\pi\tau_2}-{3\over 2}\ell\circ\ell
\right)
\Bigg)
   + \D_{\hat ac}\D_{\hat bd}\Bigg(\ell\circ\ell+ {6 \over\pi\tau_2}\Bigg) {1\over 4\pi \tau_2}
   \\
   & -\ell^I\ell^J e^L_{I\hat a}e^L_{J(d}\D_{c)\hat b} 
  \left(
   \ell\circ\ell+{18\over \pi\tau_2}
  \right)
   	 \Bigg] ,\\
&\nabla_{(a}\cdot\nabla_{b)}C = 	 \widetilde C_{\hat a\hat bii} 
 = {704\pi^4\over 161}\int_{\cal F} {d^2\tau\,\tau_2^{1\over 2}\over\Delta(\tau)} \sum_{\ell\in\Lambda} q^{\ell\circ\ell\over 2} 
 e^{-2\pi \tau_2 \ell^I e^R_{Ii}e^R_{Ji} \ell^J}
  \Bigg[ 
  \ell\circ\ell-{6\over \pi\tau_2}
  \Bigg] 
\ell^I  \ell^J e^L_{I\hat a} e^L_{J\hat b} ,
\fe
where we have used (fixing the $SO(21)$ freedom)
\ie
\D e^L_{Ja}=e^R_{Ji}e^{LI}_a\D e^R_{Ii}+{1\over 2}e^{LI}_a\D e^R_{Ii} \D e^R_{Ji}
-{1\over 2}e^R_{Jj}e^{RM}_je^{LN}_a\D e^R_{Mk}\D e^R_{Nk}
+
\dots.
\fe

After a somewhat tedious but straightforward calculation, we obtain all the differential equations in \eqref{ABCdiff},
\ie
&\sum_{e,i}\widetilde A_{abcd,eeii}=-{67\over 2}A_{abcd},\\
&\sum_{c,i} B_{ab,ccii} =-{17\over 2}B_{ab},\\
&\sum_i \widetilde{A}_{abcd,\hat {e }\hat fii} =
-2A_{\hat{a}\hat b \hat c(e}\D_{f) \hat d}-
B_{\hat{a}\hat b}\D_{\hat c(e}\D_{f)\hat d}
-{\rm trace~in~} (ef),\\
&\sum_i \widetilde B_{ab,\hat c\hat dii}={144\over 25}A_{abcd}+{71\over 25}B_{\hat a(c}\D_{d)\hat b}+2C\D_{\hat a(c}\D_{d)\hat b}-{\rm trace~in~}(cd),\\
&\sum_i \widetilde{C}_{\hat a\hat bii}={550\over 483}B_{ab},
\fe
where the hatted indices are taken to be symmetric and traceless. Together with \eqref{Cdiff}, we have thus determined all the coefficients in \eqref{ABCdiff},
\ie
&a= -{67\over2} ,~~b=  - {17\over2},~~ c= {25\over2},~~ u = -2,~~v= -1,\\
&x= {144\over25},~~y= {71\over25},~~z=2,~~w= {550\over 483}.
\fe

\paragraph{Determination of $U,V,W$}

With the above 9 coefficients determined, we now arrange them into the form \eqref{UVW} and determine $U,V,W$. 
Let us start by inspecting the trace part in $(ef)$ of \eqref{UVW},
\ie
\nabla^2 f^{(4)}_{abcd} &= (21U+V) f^{(4)}_{abcd} + W\, \delta^{ef} f^{(4)}_{ef(ab} \delta_{cd)}.
\fe
Noting that $\delta^{ef} f^{(4)}_{efab}  = {25\over6}B_{ab} +{23\over3} \delta_{ab}C$, we obtain the first three equations  in \eqref{ABCdiff},
\ie\label{UVWfinal1}
&\nabla^2 A_{abcd} = (21U+V) A_{abcd},\\
&\nabla^2 B_{ab} = (21U+V +{25\over6}W)B_{ab},\\
&\nabla^2 C = (21U+V +{23\over3}W)C.
\fe

Next, the traceless part in $(ef)$ of \eqref{UVW} can be written as
\ie\label{UVWfinal2}
&\left[ \nabla_{(e } \cdot \nabla_{f)}  - {1\over 21} \delta_{ef}  \nabla^2 \right] A_{abcd} = VA_{(e (abc} \delta_{d)f)} +{V\over2} \delta_{e(a} B_{bc} \delta_{d)f)} -\text{trace in (ef),\,(abcd)},\\
&\left[ \nabla_{(e } \cdot \nabla_{f)}  - {1\over 21} \delta_{ef}  \nabla^2 \right] B_{ab}
={6\over25} \left[ 
\left({V\over2} +{25\over6}W\right) A_{efab} 
+\left( {29\over12} V +{25\over 9} W \right)B_{(e(a}\delta_{b)f)}\right.\\
&~~~~~~~~~~~~~~~~~~~~~~~~~~~~~~~~~~~\left.
+\left( {25\over6}V+{25\over 9}W \right) \delta_{(e(a} \delta_{b)f)}C
\right]-\text{trace in (ef),\,(ab)},\\
&\left[ \nabla_{(e } \cdot \nabla_{f)}  - {1\over 21} \delta_{ef}  \nabla^2 \right] C 
= {1\over 161} \left( {25\over 6} V+ {575\over 18} W \right)B_{ef}.
\fe

Matching \eqref{UVWfinal1} and \eqref{UVWfinal2} with \eqref{ABCdiff}, we find the 9 coefficients $a,b,c,u,v,x,y,z,w$ are indeed determined by $U,V,W$, which are
\ie
U=-{3\over2},~~~~V=-2,~~~~W=6.
\fe

\subsection{Six-derivative coupling $f^{(6)}$}\label{app:f6}

In this Appendix we will show that the 6-derivative term between the 21 tensor multiplets $f^{(6)}$ defined in \eqref{f6def} satisfies the following differential equation,
\ie\label{diff62}
\nabla_{(e}\cdot \nabla_{f)} f^{(6)}_{a_1a_2,a_3a_4}
&=u_1  f^{(6)}_{a_1a_2,a_3a_4} \delta_{ef}
+u_2 \left(   f^{(6)}_{ef ,a_1a_2}\delta_{a_3a_4}  +  f^{(6)}_{ef,a_3a_4}\delta_{a_1a_2} 
 -2\,  f^{(6)}_{ef ,(\underline{a_3}(a_1} \delta_{a_2) \underline{a_4})} \right)\\
&+u_3 \left( f^{(6)}_{ea_1,fa_2}\delta_{a_3a_4} + f^{(6)}_{ea_3,fa_4}\delta_{a_1a_2}
-2  f^{(6)}_{e(\underline{a_3},f(a_1}\delta_{a_2)\underline{a_4})}\right)\\
&+u_4\left( f^{(6)}_{e(\underline{a_3},a_1a_2}\delta_{f\underline{a_4})}
+ f^{(6)}_{e(\underline{a_1},a_3a_4}\delta_{f\underline{a_2})}\right)
+u_5 \left(  f^{(6)}_{e(a_2,a_1)(a_3}\delta_{a_4)f}+f^{(6)}_{e(a_4,a_3)(a_1}\delta_{a_2)f} \right),
\fe
modulo terms of the schematic form $(f^{(4)})^2$.
In the following the symmetrization on the indices $(ef)$ is always understood if not explicitly written. We have already taken the condition \eqref{f6condition} and its consequence \eqref{uup} into account.

In the following we will use an abbreviated notation to simplify the notations, $e_{Ia}\equiv e^L_{Ia}$, $\tilde e_{Ii}= e^R_{Ii}$, and $M_{AB}\equiv \text{Im}\,\Omega_{AB}$. 

From \eqref{f6}, we can write $f^{(6)}_{a_1a_2a_3a_4}$ as
\ie
&f^{(6)}_{a_1a_2,a_3a_4} ={1\over 3} \int {\prod_{G\le H}d^6\Omega_{GH} \over \Psi_{10}(\Omega) M^{1\over2} }
 \sum_{\ell^1,\ell^2\in\Lambda} \exp\left[\, i \pi \Omega_{AB} \ell^A\circ \ell^B -2\pi M_{AB} \ell^{AI}\ell^{BJ} \tilde e_{Ii}\tilde e_{Ji}\,  \right]
 \\
 &\times
 \Big[
 32\pi^4 \epsilon_{AB}  \epsilon_{CD} \ell^{AI} \ell^{BJ} \ell^{CM} \ell^{DN} e_{Ia_1} e_{Ma_2 } e_{J(\underline{a_3}} e_{N \underline{a_4})}
+8\pi^3 M^{-1} \, M_{AB} \ell^{AI} \ell^{BJ}
(e^2)_{a_1a_2,a_3a_4,IJ}\\
&
+4\pi^2 M^{-1} ( \delta_{a_1a_2} \delta_{a_3a_4} - \delta_{a_1(a_3} \delta_{a_4) a_2})
 \Big] ,
\fe
where
\ie
(e^2)_{a_1a_2,a_3a_4,IJ}&:= -\delta_{a_3a_4}e_{Ia_1}e_{Ja_2}-\delta_{a_1a_2}e_{Ia_3}e_{Ja_4} \\
&
+{1\over2} \delta_{a_2a_3}e_{I a_4 } e_{Ja_1} 
+ {1\over2}\delta_{a_1a_4}e_{I a_2 } e_{Ja_3} 
+{1\over2} \delta_{a_2a_4}e_{I a_1 } e_{Ja_3}
+{1\over2} \delta_{a_1a_3}e_{I a_2 } e_{Ja_4}  ,
\fe
with symmetrization on the $(IJ)$ indices.

Recall that under the variation $\tilde e \rightarrow \tilde e+ \delta \tilde e$, $e_{Ia}$ transforms as, up to second order,
\ie
\delta e_{Ia} &= \tilde e_{Ii} e^J_a \, \delta\tilde e_{Ji}
+{1\over 2} \left( e^{(M}_a \delta^{N)}_I -\tilde e_{Ij} \tilde e^{(M}_j e^{N)}_a\right)
\delta\tilde e_{Mi}\delta \tilde e_{Ni},
\fe
where the $M,N$ indices are raised by $\Pi^{MN}$. We will define the tensor $G, H, E, F$ as
\ie
&\delta( e_{Ia} e_{Jb} ) = G^{Ki}_{IJ,ab} \, \delta \tilde e_{Ki} + H^{MiNj}_{IJ,ab} \, \delta \tilde e_{Mi} \delta\tilde e_{Nj},\\
&\delta( e_{I_1a_1} e_{I_2a_2}e_{I_3a_3}e_{I_4a_4} ) = E^{Ii}_{I_1I_2I_3I_4,a_1a_2a_3a_4} \, \delta \tilde e_{Ii} + F^{MiNj}_{I_1I_2I_3I_4,a_1a_2a_3a_4}\, \delta \tilde e_{Mi} \delta\tilde e_{Nj}.
\fe
They can be computed straightforwardly to be
\ie
&G^{Ki} _{IJ,ab} = \tilde e_{Ii } e^K_a e_{Jb} + \tilde e_{Ji}e^K_b e_{Ia},\\
&H^{MiNj}_{IJ,ab} = 
{1\over2} \left( e^{(M}_a \delta^{N)}_I -\tilde e_{Ik } \tilde e_k^{(M} e^{N)}_a \right) e_{Jb} \delta_{ij}
+{1\over2} \left( e^{(M}_b \delta^{N)}_J -\tilde e_{Jk } \tilde e_k^{(M} e^{N)}_b \right) e_{Ia} \delta_{ij}
+\tilde e_{Ii} \tilde e_{Jj} e^M_a e^N_b,\\
&E^{Ii}_{I_1I_2I_3I_4,a_1a_2a_3a_4} = \tilde e_{I_1i} e^I_{a_1}e_{I_2a_2} e_{I_3a_3}e_{I_4a_4} +3~\text{more},\\
&F^{MiNj}_{I_1I_2I_3I_4,a_1a_2a_3a_4} = {1\over 2} \left( e^{(M}_{a_1} \delta^{N)}_{I_1} - \tilde e_{I_1k} \tilde e^{(M}_k e^{N)}_{a_1} \right) e_{I_2a_2}e_{I_3a_3} e_{I_4a_4} \delta_{ij}+3~\text{more}\\
&~~~~~~~~~~~~~~~~~~~~+\tilde e_{I_1i}\tilde e_{I_2j} e^M_{a_1} e^N_{a_2} e_{I_3a_3} e_{I_4a_4} +5~\text{more}.
\fe

Now we can compute the variation of $f^{(6)}_{a_1a_2,a_3a_4}$ under $\tilde e_{Ii} \rightarrow \tilde e_{Ii} +\delta \tilde e_{Ii}$ up to second order,
\ie
&\delta f^{(6)}_{a_1a_2,a_3a_4}  = f^{(6)\, Ii}_{a_1a_2,a_3a_4}\delta \tilde e_{Ii} 
+ f^{(6)\, MiNj}_{a_1a_2,a_3a_4}\delta \tilde e_{Mi } \delta \tilde e_{Nj}\\
&={1\over 3}  \int {\prod_{G\le H}d^6\Omega_{GH} \over \Psi_{10}(\Omega) M^{1\over2} }
 \sum_{\ell^1,\ell^2\in\Lambda} \exp\left[\, i \pi \Omega_{AB} \ell^A\circ \ell^B -2\pi M_{AB} \ell^{AI}\ell^{BJ} \tilde e_{Ii}\tilde e_{Ji}\,  \right]\\
  &\times \Big\{\,
  \Big[
  32\pi^4  (\epsilon^2 e^4 \ell^4) _{a_1a_2,(a_3a_4)}
+8\pi^3 M^{-1} \, M_{AB} \ell^{AI} \ell^{BJ} 
(e^2)_{a_1a_2,a_3a_4,IJ}
\\
&
~~~~~~~~+4\pi^2 M^{-1}\left ( \delta_{a_1a_2} \delta_{a_3a_4} - \delta_{a_1(a_3} \delta_{a_4) a_2}\right)\Big]\times \Big[
-4\pi M_{AB} \ell^{AI} \ell^{BJ} \tilde e_{Ii} \, \delta\tilde e_{Ji}\\
&
~~~~~~~~+\left(-2\pi M_{AB} \ell^{AI}\ell^{BJ} \delta_{ij} 
+8\pi^2 M_{AB}M_{CD} \ell^{AI}\ell^{BM}\ell^{CJ}\ell^{DN} \tilde e_{Mi}\tilde e_{Nj} \right)
\delta \tilde e_{Ii}\delta\tilde e_{Jj}
\Big]\\
&+32\pi^4 \epsilon_{AB} \epsilon_{CD} \ell^{AI_1}\ell^{CI_2} \ell^{B I_3} \ell^{DI_4}
\left[ \,
E^{Ii}_{I_1I_2I_3I_4,a_1a_2(a_3a_4)} \, \delta\tilde e_{Ii}
+F^{MiNj}_{I_1I_2I_3I_4,a_1a_2(a_3a_4)}\, \delta \tilde e_{Mi}\delta\tilde e_{Nj}
\,\right]\\
&+8\pi^3 M^{-1}(M_{EF}\ell^{EI }\ell^{FJ} ) \left[
-\delta_{a_3a_4}\left( G^{Ki}_{IJ,a_1a_2}\delta\tilde e_{Ki} +H^{MiNj}_{IJ,a_1a_2} \delta\tilde e_{Mi}\delta\tilde e_{Nj}\right)+5~\text{more}
\right]\\
&+\Big[
32\pi^4 \epsilon_{AB}\epsilon_{CD} \ell^{AI_1}\ell^{CI_2}\ell^{BI_3} \ell^{DI_4} E^{Mi}_{I_1 I_2I_3I_4,a_1a_2(a_3a_4)} \\
&
+8\pi^3 M^{-1}(M_{EF}\ell^{EI}\ell^{FJ} )\left(-\delta_{a_3a_4} G^{Mi}_{IJ,a_1a_2}+
5~\text{more}
 \right)
\Big]\times\left(-4\pi M_{AB}\ell^{AK} \ell^{BN} \tilde e_{Kj}\right)\,
\delta \tilde e_{Mi}\delta\tilde e_{Nj}
  \Big\}
      \fe
where we have defined
\ie
&(\epsilon^2 e^4\ell^4)_{a_1a_2,a_3a_4} = \epsilon_{AB}  \epsilon_{CD} \ell^{AI} \ell^{BJ} \ell^{CM} \ell^{DN} e_{Ia_1} e_{Ma_2 } e_{J a_3} e_{Na_4}.
\fe
Note that $(\epsilon^2 e^4\ell^4)_{a_1a_2,(a_3a_4)} = (\epsilon^2 e^4\ell^4)_{(a_1a_2),(a_3a_4)}=(\epsilon^2 e^4\ell^4)_{(a_3a_4),(a_1a_2)}$. 


The second derivative of $f^{(6)}$ is then given by
\ie
&\sum_i \tilde f^{(6)}_{a_1a_2,a_3a_4,efii}=\sum_i \left( e_{Ie}e_{Jf} f^{(6) \,IJii}+{\delta_{ef}\over2} \tilde e_{Ii} f^{(6)\, Ii}\right)\\
&={1\over 3}  \int {\prod_{G\le H}d^6\Omega_{GH} \over \Psi_{10}(\Omega) M^{1\over2} }
 \sum_{\ell^1,\ell^2\in\Lambda} \exp\left[\, i \pi \Omega_{AB} \ell^A\circ \ell^B\, \right]\\
  &\times \Big\{\,
    \Big[
  32\pi^4  (\epsilon^2 e^4 \ell^4) _{a_1a_2,(a_3a_4)}
+8\pi^3 M^{-1} \, M_{AB} \ell^{AI} \ell^{BJ}(e^2)_{a_1a_2,a_3a_4,IJ}\\
&
~~~~~~~~+4\pi^2 M^{-1}\left ( \delta_{a_1a_2} \delta_{a_3a_4} - \delta_{a_1(a_3} \delta_{a_4) a_2}\right)\Big]\times \Big[
  \delta_{ef} M_{AB} {\partial\over \partial M_{AB}}\\
&
~~~~~~~~+\left(-10\pi M_{AB} \ell^{AI}\ell^{BJ} 
-4\pi M_{AB}M_{CD} \ell^{AI}\ell^{CJ}{\partial \over \partial M_{BD}} \right)
e_{Ie}e_{Jf}
\Big]\\
&+32\pi^4 \epsilon_{AB} \epsilon_{CD} \ell^{AI_1}\ell^{CI_2} \ell^{B I_3} \ell^{DI_4}
F^{MiNi}_{I_1I_2I_3I_4,a_1a_2(a_3a_4)}\,   e_{Me} e_{Nf}\\
&+8\pi^3 M^{-1}(M_{EF}\ell^{EI }\ell^{FJ} )  e_{Me}e_{Nf}
\left(\,- \delta_{a_3a_4}H^{MiNi}_{IJ,a_1a_2} 
+5~\text{more} \,\right)\\
&+\Big[
32\pi^4 \epsilon_{AB}\epsilon_{CD} \ell^{AI_1}\ell^{CI_2}\ell^{BI_3} \ell^{DI_4} E^{Mi}_{I_1 I_2I_3I_4,a_1a_2(a_3a_4)} \\
&
+8\pi^3 M^{-1}(M_{EF}\ell^{EI}\ell^{FJ} )\left(-\delta_{a_3a_4} G^{Mi}_{IJ,a_1a_2}+5~\text{more}\right)
\Big]\times\left(-4\pi M_{AB}\ell^{AK} \ell^{BN} \tilde e_{Ki}\right)\,
e_{M(\underline{e}}e_{N\underline{f})}  \Big\}\\
  &\times \exp\left[ -2\pi M_{AB} \ell^{AI}\ell^{BJ} \tilde e_{Ii}\tilde e_{Ji}\right]
 ,     \fe
    where we have used $E^{Ii}_{I_1I_2I_3I_4,a_1a_2(a_3a_4)} \tilde e_{Ij}=0$ and $G^{Ii}_{I_1I_2,a_1a_2} \tilde e_{Ij}=0$.

Let us now study the different powers of $\ell$ terms in the integrand. Note that since we can replace $\ell^{AI}\ell^{BJ} \tilde e_{Ii}\tilde e_{Ji}$ by $-{1\over 2\pi} {\partial \over \partial M_{AB}}$, $\tilde e_{Ii}$ should be treated as $\ell^{-1}$ in the power counting.  Also note that the tensors $G, H, E, F$ contain factors of $\tilde e_{Ii}$.

First let us note that the $\ell^6$ terms cancel as in the 4-derivative case after integration by parts. Moving on to the $\ell^4$ terms, they can be organized to be
\ie
&\sum_i \tilde f^{(6)}_{a_1a_2,a_3a_4,efii} \Big|_{\ell^4} =
{1\over 3}  \int {\prod_{G\le H}d^6\Omega_{GH} \over \Psi_{10}(\Omega) M^{1\over2} }
 \sum_{\ell^1,\ell^2\in\Lambda} \exp\left[\, i \pi \Omega_{AB} \ell^A\circ \ell^B\, \right]\\
  &\times \Big\{\,
  -64\pi^4 \delta_{ef}  (\epsilon^2 e^4 \ell^4) _{a_1a_2,(a_3a_4)}\\
  &
  +32\pi^4\left[ \,  \, \delta_{a_3a_4} (\epsilon^2e^4\ell^4)_{ef,(a_1a_2)}
 +\delta_{a_1a_2} (\epsilon^2e^4\ell^4)_{ef,(a_3a_4)}
 -{1\over2}\delta_{a_2 a_3} (\epsilon^2 e^4\ell^4)_{ef,(a_1 a_4)}\right.\\
 &\left.
  -{1\over2}\delta_{a_1 a_4} (\epsilon^2 e^4\ell^4)_{ef,(a_2 a_3)}
  -{1\over2}\delta_{a_2 a_4} (\epsilon^2 e^4\ell^4)_{ef,(a_1 a_3)}
    -{1\over2}\delta_{a_1 a_3} (\epsilon^2 e^4\ell^4)_{ef,(a_2 a_4)}
 \right]
\\
 &
+ 16\pi^4 \left[ \, \delta_{a_1e}\, (\epsilon^2e^4\ell^4)_{fa_2,(a_3a_4)}
+\delta_{a_2e} \,(\epsilon^2e^4\ell^4)_{a_1 f,(a_3a_4)}
+\delta_{a_3e} \,(\epsilon^2e^4\ell^4)_{a_1 a_2,(fa_4)}
+\delta_{a_4e} \,(\epsilon^2e^4\ell^4)_{a_1 a_2,(a_3f)}\,
\right]\Big\}.
  \fe
This already fixes $u_i$'s to be
\ie\label{ui}
u_1=-2,~~u_2=1,~~u_3=0,~~u_4=1,~~u_5=0.
\fe

In the following we will show that the terms with $\ell^2$ and $\ell^0$ in the integrand also satisfies the same differential equation \eqref{diff62} with the same values of $u_i$'s. Let us start with the $\ell^0$ term in the covariant Hessian (LHS of \eqref{diff62}), 
\ie
&\sum_i \tilde f^{(6)}_{a_1a_2,a_3a_4,efii} \Big|_{\ell^0}\propto
{1\over 3}  \int {\prod_{G\le H}d^6\Omega_{GH} \over \Psi_{10}(\Omega) M^{1\over2} }
 \sum_{\ell^1,\ell^2\in\Lambda} \exp\left[\, i \pi \Omega_{AB} \ell^A\circ \ell^B\, \right]\\
 &\times
M_{AB}{\partial\over \partial M_{AB}} \exp\left[
-2\pi M_{AB}\ell^{AI}\ell^{BJ} \tilde e_{Ii}\tilde e_{Ji}
\right]=0.
  \fe
Hence we need to show that the righthand side of \eqref{diff62} is also zero when replacing $f^{(6)}_{a_1a_2,a_3a_4}$ by its $\ell^0$ term in the integrand, namely, $f^{(6)}_{a_1a_2,a_3a_4}\rightarrow (\delta_{a_1a_2} \delta_{a_3a_4} -\delta_{a_1(a_3} \delta_{a_4)a_2} )$. Indeed, under this replacement the righthand side of \eqref{diff62} is zero with $u_i$'s given by \eqref{ui}
\ie
&u_1 \left(\delta_{a_1a_2} \delta_{a_3a_4} -\delta_{a_1(a_3} \delta_{a_4)a_2} \right) \delta_{ef}\\
&
+2u_2 \left( \delta_{ef}\delta_{a_1a_2} \delta_{a_3a_4} -\delta_{e(a_1} \delta_{a_2)f} \delta_{a_3a_4} 
-\delta_{ef} \delta_{(a_3 (a_1}\delta_{a_2)a_4)} +\delta_{e(a_1 }\delta_{a_2)(a_3} \delta_{a_4)f}\right)\\
&
+2u_4\left( \delta_{ea_3 } \delta_{a_4)f} \delta_{a_1a_2} -\delta_{e(a_1}\delta_{a_2)(a_3} \delta_{a_4)f}\right)\\
&+(a_1\leftrightarrow a_3,~a_2\leftrightarrow a_4)=0.
\fe
Again, the symmetrization on the indices $(ef)$ is implicitly understood.

Next, the $\ell^2$ terms can be organized as
\ie\label{f6l2}
&\sum_i \tilde f^{(6)}_{a_1a_2,a_3a_4,efii} \Big|_{\ell^2} =
{1\over 3}  \int {\prod_{G\le H}d^6\Omega_{GH} \over \Psi_{10}(\Omega) M^{1\over2} }
 \sum_{\ell^1,\ell^2\in\Lambda} \exp\left[\, i \pi \Omega_{AB} \ell^A\circ \ell^B -2\pi M_{AB} \ell^{AI}\ell^{BJ} \tilde e_{Ii}\tilde e_{Ji}\, \right]\\
  &\times e_{Jf} \Big\{
  \Big[
- 8\pi^3 \,  \delta_{ef} (e^2)_{a_1a_2,a_3a_4,IJ}
-16\pi^3  \, \left ( \delta_{a_1a_2} \delta_{a_3a_4} - \delta_{a_1(a_3} \delta_{a_4) a_2}\right) e_{Ie}e_{Jf}
\Big]\\
&-4\pi^3\Big(
2 \delta_{a_1e} \delta_{a_2f} e_{Ia_3}e_{Ja_4}
 +2 \delta_{a_3e} \delta_{a_4f} e_{Ia_1}e_{Ja_2}
\\
& -\delta_{a_1e} \delta_{a_3f} e_{Ia_2}e_{Ja_4}
 -\delta_{a_1e} \delta_{a_4f} e_{Ia_2}e_{Ja_3} 
 -\delta_{a_2e} \delta_{a_3f} e_{Ia_1}e_{Ja_4}
  -\delta_{a_2e} \delta_{a_4f} e_{Ia_1}e_{Ja_3}\Big)
\\
&-4\pi^3 \Big[
- \delta_{a_3a_4} (\delta_{a_1e}e_{Ia_2} +\delta_{a_2e} e_{Ia_1})
- \delta_{a_1a_2} (\delta_{a_3e}e_{Ia_4} +\delta_{a_4e} e_{Ia_3})\\
&
+{1\over2} \delta_{a_2a_3} (\delta_{a_1e}e_{Ia_4} +\delta_{a_4e} e_{Ia_1})+3~\text{more}
\Big]
 \Big\}.
 \fe
We need to match the second derivative of $f^{(6)}$ given above with the righthand side of \eqref{diff62} at the $\ell^2$ order in the integrand. For example, the coefficient for $\delta_{ef} (e^2)_{a_1a_2,a_3a_4,IJ}$ on the righthand side of \eqref{diff62} is $8\pi^3(u_1+u_2)=-8\pi^3$, which agrees with the coefficient the second derivative $\tilde f^{(6)}$. Similarly one can show that the $\ell^2$ terms agree on both sides of \eqref{diff62}.

In conclusion, we have checked that $f^{(6)}_{a_1a_2,a_3a_4}$ given in \eqref{f6} satisfies the following differential equation,
\ie
2\nabla_{(e}\cdot \nabla_{f)} f^{(6)}_{a_1a_2,a_3a_4}
&=-2  f^{(6)}_{a_1a_2,a_3a_4} \delta_{ef}
+ \left(   f^{(6)}_{ef ,a_1a_2}\delta_{a_3a_4}  +  f^{(6)}_{ef,a_3a_4}\delta_{a_1a_2} 
 -2\,  f^{(6)}_{ef ,(\underline{a_3}(a_1} \delta_{a_2) \underline{a_4})} \right)\\
&+\left( f^{(6)}_{e(\underline{a_3},a_1a_2}\delta_{f\underline{a_4})}
+f^{(6)}_{e(\underline{a_1},a_3a_4}\delta_{f\underline{a_2})} \right)+(e\leftrightarrow f)\,,
\fe
modulo the $(f^{(4)})^2$ term that is determined in Section \ref{sec:example} and Appendix \ref{app:5dMSYM}.

\section{Relation to 5$d$ MSYM amplitudes}\label{app:5dMSYM}

In Section \ref{sec:example}, we discuss how the numerical coefficients $v_1,v_2,v_3$ for the $(f^{(4)})^2$ term in \eqref{f6diffeqn:gen} can be fixed from the 6$d$ $(2,0)$ SCFT limit, where a similar differential equation holds \cite{CDLY}. The four-point 4- and 6-derivative couplings on the tensor branch of the 6$d$ $(2,0)$ SCFT can be in turn computed by the one- and two-loop amplitudes in 5$d$ maximal SYM on its Coulomb branch\cite{Cordova:2015vwa}. Therefore, to determine these coefficients, we will fix the relative normalization between the $F^4$ and $D^2F^4$ couplings in the Coulomb branch effective action of 5$d$ maximal SYM and the $T^5$ compactified heterotic string amplitudes in this appendix.

\subsection{Four-derivative coupling $f^{(4)}$}\label{app:f4sym}
In this subsection, we would like to fix the relative normalization between the $F^4$ coupling  from one-loop heterotic string amplitude and that from one-loop 5$d$ maximal SYM on its Coulomb branch by looking at a point of enhanced ADE gauge symmetry in the heterotic moduli space and a degeneration limit of the genus one Riemann surface. A similar reduction of the genus one and two amplitudes in the type II string theory to supergravity amplitudes was considered in \cite{Tourkine:2013rda}.

\begin{figure}[h]
\centering
\includegraphics[scale=0.95]{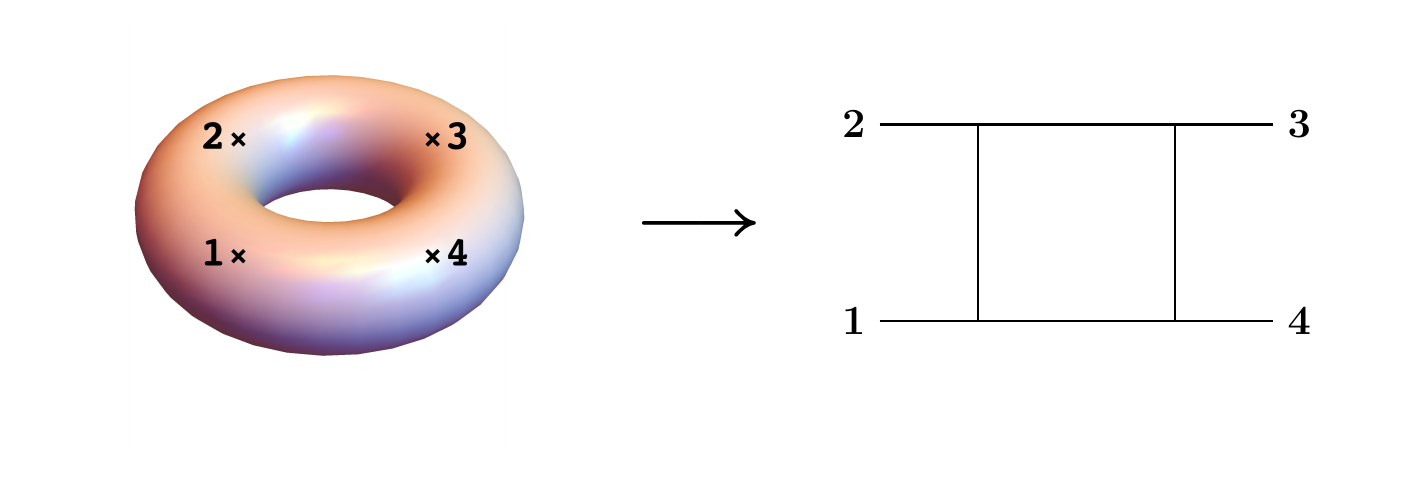}
\caption{The reduction of the genus one $T^5$ compactified heterotic string amplitude $\mathcal{A}_1|_{F^4}$ to the one-loop amplitude $\mathcal{A}_1^{SYM}$ in 5$d$ maximal SYM. }
\end{figure}

Recall that the heterotic one-loop amplitude is
\ie
\left.{\cal A}_1\right|_{F^4} 
&=  \left. {\partial^4\over  \partial y^{a_1}\cdots\partial y^{a_4}} \right|_{y=0} \int_{\cal F}  {d^2\tau\over \tau_2^2} {\tau_2^{5\over 2} \Theta_\Lambda(y|\tau,\bar\tau) \over  \Delta(\tau)},
\fe
with the theta function $\Theta_\Lambda$ defined by
\ie
\Theta_\Lambda(y|\tau,\bar\tau) &= e^{{\pi \over 2\tau_2 }y\circ y } \sum_{\ell\in \Lambda} e^{\pi i \tau \ell_L^2 - \pi i \bar\tau \ell_R^2 + 2\pi i \ell\circ y}
\\
&=  e^{{\pi \over 2\tau_2 }y\circ y } \sum_{\ell\in \Lambda} e^{\pi i \tau \ell\circ\ell - 2 \pi \tau_2 \ell_R^2 + 2\pi i \ell\circ y}.
\fe
Let us inspect the contributions to the integral in the large $\tau_2$ regime, where $\Delta(\tau)$ can be approximated by $q=e^{2\pi i\tau}$.  Then $\Theta_\Lambda$ is dominated by the contribution from $\ell\circ\ell=\ell_L^2-\ell_R^2=2$, and we have
\ie
\left. {\cal A}_1\right|_{F^4} &\rightarrow   (2\pi)^4 \int  d\tau_2 \, \tau_2^{1\over 2}   \sum_{\ell\circ\ell=2} \ell^L_{a_1} \ell^L_{a_2} \ell^L_{a_3} \ell^L_{a_4}
e^{-2\pi \tau_2 \ell_R^2  }  .
\fe
In the limit of the moduli space where $\ell_R\to 0$ for some of the $\ell\circ\ell=2$ lattice vectors, the dominant contribution takes the form of the one-loop contribution from integrating out $W$-bosons labeled the root vectors $\ell$ in 5$d$ maximal SYM. Here $\ell_R^2$ is proportional to the $W$-boson mass squared, and $\ell^L_a$ labels the charge of the $W$-boson with respect to the $a$-th Cartan generator.

To compare the normalization with the 5$d$ SYM one-loop amplitude, we use the Schwinger parametrization to write down the contribution from the diagrams involving light internal $W$-bosons, which are labeled by the root vectors $\ell^L$,
\ie
\cA_1^{SYM} &= \sum_{(\ell^L)^2 = 2} 3\int dt {t^3\over 3!}\ell^L_{a_1} \ell^L_{a_2} \ell^L_{a_3} \ell^L_{a_4}
\int {d^5p\over {(2\pi )^5}}e^{-t(p^2+m^2)}
\\
&= {1\over 2^6 \pi^{5\over 2}}\int dt~ t^{1\over 2} \sum_{(\ell^L)^2 = 2} \ell^L_{a_1} \ell^L_{a_2} \ell^L_{a_3} \ell^L_{a_4}
 e^{-t  m^2 }
\fe
Identifying $m^2=2\pi \ell_R^2$, we fix the relative normalization to be
\ie
\left. {\cal A}_1\right|_{F^4} &\rightarrow  2^{10}\pi^{13\over 2} \cA_1^{SYM}.
\fe

\subsection{Six-derivative coupling $f^{(6)}$}\label{app:f6sym}
In this subsection, we would like to fix the relative normalization between the $D^2F^4$ coupling  from two-loop heterotic string amplitude and that from two-loop 5$d$ maximal SYM on its Coulomb branch by looking at a point of enhanced ADE gauge symmetry in the heterotic moduli space and a degeneration limit of the genus two Riemann surface.

\begin{figure}[h]
\centering
\includegraphics[scale=0.65]{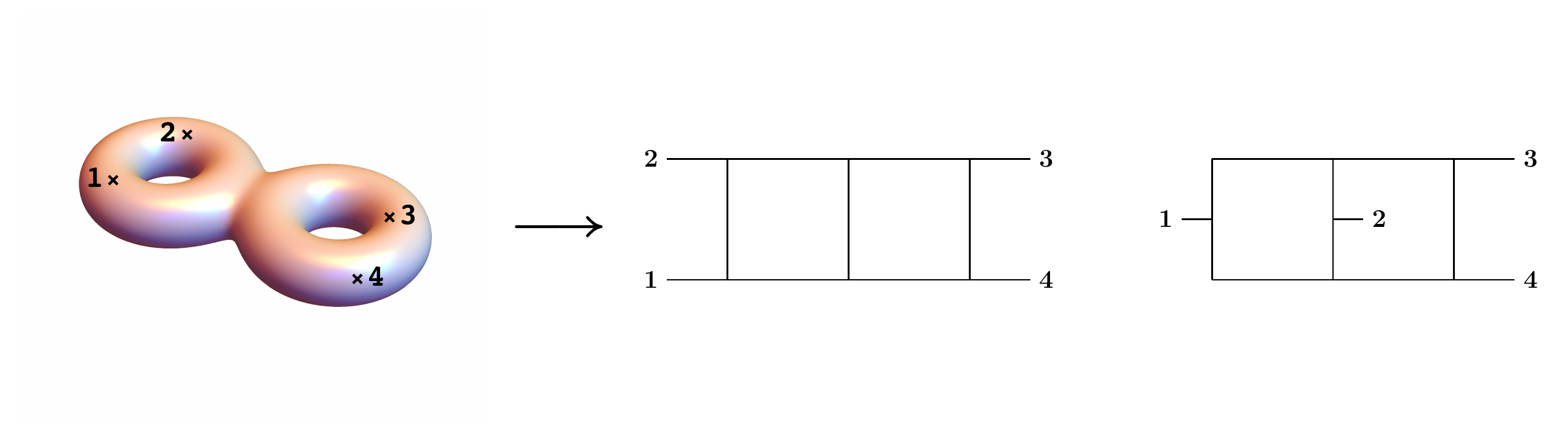}
\caption{The reduction of the genus two $T^5$ compactified heterotic string amplitude $\mathcal{A}_2|_{D^2F^4}$ to two-loop amplitudes $\mathcal{A}_2^{SYM}$ in 5$d$ maximal SYM. }\label{fig:twoloop}
\end{figure}

Recall that the heterotic two-loop amplitude is  
\ie
\left. {\cal A}_2\right|_{D^2F^4} &= \left[ {t-u\over 3} \epsilon_{IJ} \epsilon_{KL}\left. {\partial^4\over \partial y_I^{a_1} \partial y_J^{a_2} \partial y_K^{a_3} \partial y_L^{a_4}} \right|_{y=0} + (2~{\rm perms}) \right] \int_{{\cal F}_2} {\prod_{I\leq J} d^2\Omega_{IJ}\over (\det{\rm Im}\Omega)^{1\over 2} \Psi_{10}(\Omega)}\Theta_\Lambda(y|\Omega,\bar\Omega),
\fe
with the theta function given by
\ie
\Theta_\Lambda (y|\Omega,\bar\Omega)& \equiv \sum_{\ell^1, \ell^2 \in\Lambda} e^{\pi i \Omega_{AB} \ell^A_L \cdot\ell^B_L - \pi i \bar\Omega_{AB}\ell^A_R\cdot\ell^B_R + 2\pi i \ell^A\circ y_A + {\pi\over 2}(({\rm Im}\Omega)^{-1})^{AB} y_A \cdot y_B }\\
&=\sum_{\ell^1, \ell^2 \in\Lambda} e^{i\pi  \Omega_{AB} \ell^A \circ\ell^B -2\pi \text{Im}\, \Omega_{AB}\ell^{A}_R\cdot \ell^{B}_R+ 2\pi i \ell^A\circ y_A + {\pi\over 2}(({\rm Im}\Omega)^{-1})^{AB} y_A \cdot y_B }.
\fe
Each component of ${\rm Re}\Omega_{AB}$ has periodicity 1. The imaginary part of the period matrix can be written as
\ie
{\rm Im}\Omega = \begin{pmatrix} t_1 + t_3 & t_3 \\ t_3 & t_2+t_3 \end{pmatrix},
\fe
with $\det{\rm Im}\Omega = t_1t_2 + t_1t_3+t_2t_3$. In the limit of large positive $t_1,t_2,t_3$, this corresponds to the genus two Riemann surface degenerating into three long tubes, of length $t_1, t_2, t_3$ respectively. We can also write
\ie
& {\rm Im}\Omega_{AB} \ell^A\cdot \ell^B = t_1 (\ell^1)^2 + t_2 (\ell^2)^2 + t_3 (\ell^1+\ell^2)^2,
\\
& (({\rm Im}\Omega)^{-1})^{AB} y_A\cdot y_B = {t_1 y_2^2 + t_2 y_1^2 + t_3 (y_1 - y_2)^2\over t_1t_2+t_1t_3+t_2t_3}.
\fe
In the limit of large positive $t_1, t_2, t_3$, the theta function, apart from the term 1 which vanishes upon taking $y$-derivative, is dominated by the terms involving lattice vectors $\ell$ such that $\ell_L^2+\ell_R^2$ is close to 2, when the lattice embedding is near an ADE point in the moduli space. The Igusa cusp form $\Psi_{10}(\Omega)$, on the other hand, behaves as 
\ie
\Psi_{10}(\Omega) \to e^{2\pi i {\rm Re}(\Omega_{11}+\Omega_{22}-\Omega_{12})} e^{-2\pi (t_1+t_2+t_3)},
\fe
where we have used the product expression for $\Psi_{10}(\Omega)$,
\ie
\Psi_{10}(\Omega) = e^{2\pi i (\tau +\sigma+\nu)} \prod_{(n,k,\ell)>0} \left(
1- e^{2\pi i(n \tau +k \sigma +\ell \nu)}
\right).
\fe
Here $(n,k,\ell)>0$ means that $n,k\ge 0$, $\ell\in \mathbb{Z}$, and in the case when $n=k=0$, the product is only over $\ell<0$. In the above expression we parametrize $\Omega$ as
\ie
\Omega= \left(\begin{array}{cc}\tau & \nu \\\nu & \sigma\end{array}\right) .
\fe 
The integration over ${\rm Re}\Omega_{AB}$ then picks out the terms in the theta function with
\ie
\ell^1\circ\ell^1= \ell^2\circ\ell^2=(\ell^1+\ell^2)^2= 2,
\fe
giving the factor
\ie
\exp\left[-2\pi (t_1 (\ell^1_R)^2+t_2 (\ell^2_R)^2+t_3 (\ell^1_R+\ell^2_R)^2) +2\pi i \ell^A\circ y_A 
\right].
\fe
We are interested in the limit where $(\ell_R^1)^2$, $(\ell_R^2)^2$, and $(\ell_R^1+\ell_R^2)^2$ are small, and correspond to $W$-boson masses of three propagators in the two-loop diagram. We have (in the rest of this section we will not distinguish $\ell^I_a$ with $(\ell_L)^I_a$ since in the limit of interest $(\ell_R)^I_a \to 0$)
\ie\label{alim}
\left. {\cal A}_2\right|_{D^2F^4} &\to {t-u\over 3}(2\pi)^4\sum_{(\ell^1)^2=(\ell^2)^2=(\ell^1+\ell^2)^2=2}  \epsilon_{IJ}\epsilon_{KL} \ell^I_{a_1} \ell^J_{a_2} \ell^K_{a_3}\ell^L_{a_4} 
\\
&~~~~\times \int {dt_1 dt_2 dt_3\over (t_1t_2+t_1t_3+t_2t_3)^{1\over 2}} e^{-2\pi (t_1 (\ell^1_R)^2+t_2 (\ell^2_R)^2+t_3 (\ell^1_R+\ell^2_R)^2)} + ({\rm \text{cyclic perms in } 2,3,4}).
\fe
Here $\ell^I_a$ is the eigenvalue of the Cartan generator $T_a$ on the $W$-boson labeled by the root vector $\ell^I$, on the propagator of length $t_I$, $I=1,2$. On the third propagator of length $t_3$, the $W$-boson has charge $\ell^1_a+\ell^2_a$ with respect to $T_a$.

Let us compare this with the two-loop amplitude at 6-derivative order in 5$d$ SYM, whose contribution from the diagrams involving two light internal $W$-bosons 
takes the form
\ie
\cA^{SYM}_2 &= {s \over 2} \sum_{(\ell^{1L})^2=(\ell^{2L})^2=(\ell^{1L}+\ell^{2L})^2=2} \int dt_1 dt_2 dt_3
\\
& \quad \Bigg[ {t_1^2t_2^2} \ell^1_{a_1}\ell^1_{a_2}\ell^2_{a_3}\ell^2_{a_4} +\text{5 more} \\
& \hspace{.5in} - {t_1^2 t_2t_3} \ell^1_{a_1}\ell^1_{a_2}(-\ell^1_{a_3}-\ell^2_{a_3})\ell^2_{a_4} 
-{t_1^2 t_2t_3} ( -\ell^1_{a_1}-\ell^2_{a_1})\ell^2_{a_2}\ell^1_{a_3}\ell^1_{a_4} 
+\text{10 more} \Bigg] 
\\
& \quad \times \int {d^5p_1 d^5p_2\over (2\pi)^{10}} e^{-\sum_{i=1}^3 t_i(p_i^2+m_i^2)} +({\rm \text{cyclic perms in } 2,3,4}),
\fe
where the first and the second lines come from the first and the second two-loop diagrams in  Figure \ref{fig:twoloop}, respectively. 
The term proportional to $t_1^2t_2^2$, for instance, comes from the two-loop diagram with two external lines (with Cartan label $a_1, a_2$) attached to the propagator of length $t_1$ and two external lines (with Cartan label $a_3, a_4$) attached to the propagator of length $t_2$. The $\cdots$ stand for all the other possible assignments of the $W$-boson root vectors $\ell^1,\,\ell^2,\, -\ell^1-\ell^2$ to each internal propagator. 

We can identify $m_1^2=2\pi(\ell^1_R)^2$, $m_2^2=2\pi(\ell^2_R)^2$, $m_3^2=2\pi(\ell^1_R+\ell^2_R)^2$.
The factor in the bracket, after multiplication by $s$ and summation over permutations, can be organized into the form (taking into account $s+t+u=0$)  
\ie
&s {(t_1t_2+t_1t_3+t_2t_3)^2} \left(\ell^1_{a_1} \ell^1_{a_2}\ell^2_{a_3}\ell^2_{a_4}+\ell^2_{a_1} \ell^2_{a_2}\ell^1_{a_3}\ell^1_{a_4} \right) + ({\rm \text{cyclic perms in } 2,3,4})
\\
&={2\over 3} s {(t_1t_2+t_1t_3+t_2t_3)^2} \left(\ell^1_{a_1} \ell^1_{a_2}\ell^2_{a_3}\ell^2_{a_4}+\ell^2_{a_1} \ell^2_{a_2}\ell^1_{a_3}\ell^1_{a_4}- 2\ell^1_{(a_1} \ell^2_{a_2)}\ell^1_{(a_3}\ell^2_{a_4)} \right) +({\rm \text{cyclic perms in } 2,3,4})
\\
&= {s\over 3} {(t_1t_2+t_1t_3+t_2t_3)^2} (\epsilon_{IK}\epsilon_{JL}+\epsilon_{IL}\epsilon_{JK} ) \ell^I_{a_1} \ell^J_{a_2}\ell^K_{a_3}\ell^L_{a_4}+ ({\rm \text{cyclic perms in } 2,3,4}).
\fe
Notice that only terms with two $\ell^1$ and two $\ell^2$ will survive after summing over the $s,t,u$ channels. 
Hence the SYM two-loop amplitude can be put into the form
\ie
 \cA^{SYM}_2
&= 2^{-11}\pi^{-5} \sum_{(\ell^{1L})^2=(\ell^{2L})^2=(\ell^{1L}+\ell^{2L})^2=2} 
\\
& \quad \left[ {s\over 3}  (\epsilon_{IK}\epsilon_{JL}+\epsilon_{IL}\epsilon_{JK} ) \ell^I_{a_1} \ell^J_{a_2}\ell^K_{a_3}\ell^L_{a_4}+ ({\rm \text{cyclic perms in } 2,3,4})\right] \\
& \quad\quad \times \int {dt_1 dt_2 dt_3\over (t_1t_2+t_1t_3+t_2t_3)^{1\over 2}} e^{-\sum_i t_i m_i^2}.
\fe
This is indeed proportional to (\ref{alim}),
\ie
\left. {\cal A}_2\right|_{D^2F^4} &\to 2^{15} \pi^9\cA^{SYM}_2.
\fe

\bibliographystyle{JHEP}
\bibliography{20sugrarefs}

 \end{document}